\documentclass[usenatbib]{mn2e}
\usepackage{graphicx, txfonts, natbib}
\usepackage{epsf}
\usepackage{amssymb}
\usepackage{epstopdf}
\usepackage{longtable}
\usepackage{subfigure}
\usepackage{comment}
\newcommand{\bv}{\mbox{$B\!-\!V$}}

\newcommand{\msun}{\mbox{M$_{\odot}$}}

\newcommand{\kms}{\mbox{$\rm{km}\,s^{-1}$}}

\newcommand{\nick}{\mbox{$^{56}$Ni}}

\DeclareMathAlphabet{\mathsc}{OT1}{cmr}{m}{sc}
\def\testbx{bx}%
\DeclareRobustCommand{\ion}[2]{%
\relax\ifmmode
\ifx\testbx\f@series
{\mathbf{#1\,\mathsc{#2}}}\else
{\mathrm{#1\,\mathsc{#2}}}\fi
\else\textup{#1\,{\mdseries\textsc{#2}}}%
\fi}

\newcommand{\Caii} {\ion{Ca}{ii}}

\newcommand{\SiII} {\ion{Si}{ii}}

\newcommand{\Mgii} {\ion{Mg}{ii}}

\newcommand{\SiiNF}{\ion{S}{ii}}
\newcommand{\Coii}{\ion{Co}{ii}}

\begin{document}
\title[NUV observations of SNe Ia] {\textit{Hubble Space Telescope} studies of low-redshift Type Ia supernovae: Evolution with redshift and ultraviolet spectral trends}
 \author[K. Maguire et al.]
  {K.~Maguire,$^1$\thanks{E-mail: kate.maguire@astro.ox.ac.uk}
   M.~Sullivan,$^1$ R.~S.~Ellis,$^2$ P.~E.~Nugent,$^{3,4}$ D.~A.~Howell,$^{5,6}$ A.~Gal-Yam,$^7$
     \newauthor  J.~Cooke,$^8$,   P.~Mazzali,$^{9,10}$   Y-C.~Pan,$^1$ B.~Dilday,$^{5,6}$ R.~C.~Thomas,$^{3}$  I.~Arcavi,$^{7}$ S.~Ben-Ami,$^7$
     \newauthor  D.~Bersier,$^{11}$ F.~B.~Bianco,$^{5,6}$ B.~J.~Fulton,$^{5}$ I.~Hook,$^{1,12}$ A.~Horesh,$^{2}$ E.~Hsiao,$^{13}$ P.~A.~James,$^{11}$    
     \newauthor   P.~Podsiadlowski,$^1$ E.~S.~Walker,$^{14}$ O.~Yaron,$^{7}$ M.~M.~Kasliwal,$^{15}$  R.~R.~Laher,$^{16}$ N.~M.~Law,$^{17}$
     \newauthor  E.~O.~Ofek,$^7$ D.~Poznanski$^{18}$ J.~Surace$^2$\\
      $^1$Department of Physics (Astrophysics), University of Oxford, DWB, Keble Road, Oxford OX1 3RH, UK\\
      $^2$Cahill Center for Astrophysics, California Institute of Technology, East California Boulevard, Pasadena, CA 91125, USA\\
      $^3$Computational Cosmology Center, Lawrence Berkeley National Laboratory, 1 Cyclotron Road, Berkeley, CA 94720, USA\\
      $^4$Department of Astronomy, University of California, Berkeley, CA 94720-3411, USA\\
      $^5$Las Cumbres Observatory Global Telescope Network, Goleta, CA 93117, USA\\
      $^6$Department of Physics, University of California, Santa Barbara, CA 93106-9530, USA\\
      $^7$Astrophysics Group, Weizmann Institute of Science, Rehovot 76100, Israel\\
      $^8$Swinburne University of Technology, Victoria 3122, Australia\\    
      $^{9}$INAF-Osservatorio Astronomico, vicolo dell'Osservatorio, 5, I-35122 Padova, Italy \\
      $^{10}$Max-Planck Institut f\"ur Astrophysik, Karl-Schwarzschild-Str. 1, D-85748 Garching, Germany \\
      $^{11}$Astrophysics Research Institute, Liverpool John Moores University, Twelve Quays House, Egerton Wharf, Birkenhead CH41 1LD, UK\\
      $^{12}$INAF-Osservatorio Astronomico di Roma, via Frascati, 33, 00040 Monte Porzio Catone (Roma), Italy\\
      $^{13}$Carnegie Observatories,  Las Campanas Observatory,Casilla 601, La Serena, Chile\\
      $^{14}$Scuola Normale Superiore di Pisa, Piazza dei Cavalieri 7, 56126 Pisa, Italy\\
      $^{15}$Observatories of the Carnegie Institution for Science, 813 Santa Barbara St., Pasadena CA 91101, USA\\
      $^{16}$Spitzer Science Center, California Institute of Technology,  M/S 314-6, Pasadena, CA 91125, USA\\
        $^{17}$Dunlap Institute for Astronomy and Astrophysics, University of Toronto, 50 St. George Street, Toronto M5S 3H4, Ontario, Canada\\
       $^{18}$School of Physics and Astronomy, Tel Aviv University, Tel Aviv 69978, Israel}

\maketitle

\begin{abstract}
 We present an analysis of the maximum light, near ultraviolet (NUV;
  $2900<\lambda<5500$ \AA) spectra of 32 low redshift ($0.001<z<0.08$)
  Type Ia supernovae (SNe Ia), obtained with the \textit{Hubble Space
    Telescope (HST)} using the Space Telescope Imaging Spectrograph.
  We combine this spectroscopic sample with high-quality $gri$ light
  curves obtained with robotic telescopes to measure SN Ia photometric
  parameters, such as stretch (light curve width), optical colour, and
  brightness (Hubble residual). 
   By comparing our new data to a comparable sample of SNe Ia at intermediate redshift ($0.4<z<0.9$), we detect modest spectral evolution (3-$\sigma$), in the sense that our mean low redshift NUV spectrum has a depressed flux compared to its intermediate redshift counterpart. We also see a strongly increased dispersion about the
  mean with decreasing wavelength, confirming the results of earlier
  surveys. We show these trends are consistent with changes in metallicity as predicted by contemporary SN Ia spectral models.
   We also examine the properties of various NUV spectral diagnostics in
  the individual SN spectra. We find a general correlation between SN
  stretch and the velocity (or position) of many NUV spectral
  features.  In particular, we observe that higher stretch SNe have larger \Caii\ H\&K velocities, which also correlate with host galaxy stellar mass. This latter trend is probably driven by the well-established correlation between stretch and host galaxy stellar mass.  We find no significant trends
  between UV spectral features and optical colour.  
  Mean spectra constructed according to
  whether the SN has a positive or negative Hubble residual show very
  little difference at NUV wavelengths, indicating that the NUV evolution and
  variation we identify does not directly correlate with Hubble diagram residuals.
  Our work confirms and strengthens earlier conclusions regarding the complex behaviour of SNe Ia in the NUV spectral region, but suggests the correlations we find are more useful in constraining progenitor models rather than improving the use of SNe\,Ia as cosmological probes.  
\end{abstract}


\section{Introduction} \label{intro}

The use of Type Ia supernovae (SNe Ia) as cosmological probes is now
well established; significant advancements in constraining the
cosmological parameters using distant SNe Ia \citep{rie07, kes09,
  sul11} have been made since the initial discovery of the
accelerating Universe at the end of the last century \citep{rie98,
  per99}. SNe Ia are believed to result from the thermonuclear
explosion of an accreting carbon-oxygen (CO) white dwarf in a binary system. Although the nature of the
primary exploding star has recently been confirmed as a white dwarf
\citep{nug11,blo12}, the nature of the secondary, mass-donating
companion star remains a mystery.  Generally, two possibilities are
considered -- double-degenerate systems (with two white dwarfs) and
single degenerate scenarios (a white dwarf plus a giant, sub-giant or
main sequence star).  Recent results have indicated a diversity in the
form of this companion star from SN event to SN event, with evidence
that some SNe Ia have red giant companions \citep{dil12}, evidence
that directly excludes red-giants in other cases \citep[][]{nug11, li11, cho12, hor12, mar12, rus12},
evidence that favours single degenerate systems \citep{ste11}, and
evidence that favours double-degenerates \citep{sch12}. If SNe Ia do
indeed result from two (or more) progenitor channels, then
understanding their differences and their dependence on host galaxy
properties (or stellar populations) is of utmost importance for their
use as distance indicators.

The metallicity or composition of the exploding white dwarf is also likely
to affect the photometric properties of SNe Ia \citep{tim03,kas09}, an
effect that may have been detected observationally: Hubble residuals
correlate with the stellar mass of the SN Ia host galaxies
\citep{kel10,sul10,lam10}, which can be interpreted as a crude proxy for
metallicity. Gas-phase metallicity measurements have also been found to correlate with Hubble residuals \citep{dan11}. SN Ia spectroscopy is a potential probe of this
astrophysics, and the rest-frame ultraviolet (UV) in particular is
expected to show strong signatures of metallicity/compositional
effects \citep[e.g.,][]{hof98, len00, sau08, wal12}. The NUV spectral region is dominated by numerous overlapping lines, that make identifying individual features difficult and results in heavy line blanketing. The UV flux of SNe Ia at early times is formed mainly through the process
of `reverse-fluorescence', the scattering of photons from longer to
shorter wavelengths \citep{luc99, maz00}.  The UV photons emitted
through this process are predominantly found to originate from a thin
layer of material at high-velocity \citep{maz00}. Therefore, the NUV
spectral region is a good probe of the outer layers of the SN ejecta.

Generally, spectral studies have taken two different approaches:
either a comparison of mean spectra, testing for evolutionary effects
in composition or velocity profile with redshift, or measurements of
individual spectral features, testing for relationships between these
spectral features and SN photometric properties.  Some spectral comparisons
between low- and high-redshift ($z$) samples have suggested that there may be some evolution with redshift in the mean NUV spectrum \citep[][hereafter, E08]{fol08a, bal09, sul09, fol12b, ell08}.  However, due to the difficulties in
obtaining UV spectra of low-$z$ SNe Ia, studies of the
metallicity-influenced UV region have been limited by a small number
of spectra of just a handful of SNe Ia at low redshift, and have not
been able to probe either the full range of ``normal'' SN Ia
properties or host galaxy properties.  Observations at higher
redshift -- where the rest-frame near-UV (NUV) is redshifted into the
optical and observations are dominated by UV flux -- have also shown an increased scatter at NUV wavelengths compared
to the optical (E08). The limited \textit{Hubble Space Telescope
  (HST)} and \textit{SWIFT} spectroscopy of low-$z$ events that is available supports the higher-$z$ studies \citep{buf09, coo11, wan11}, as does \textit{Swift} UV photometry
which illustrates an increased dispersion in UV colours implying
larger spectral variations \citep{mil10,bro10}.  One interpretation is
that the larger variations at these wavelengths are due to differing
progenitor metallicity and composition; an increased dispersion is in
broad agreement with the results of SN Ia UV spectral modelling
\citep{wal12}.

Studies of particular SN Ia spectral features have usually
concentrated on improving their use as distance indicators, either by
measuring spectral feature equivalent widths, velocities, or spectral
flux ratios.  Various empirical trends of differing statistical
significance have been claimed between SN light-curve width and
spectral feature equivalent width \citep{bro08, wal11, cho11}, between SN luminosity
and spectral flux ratios \citep{fol08b,bai09,blo11, sil12}, and between SN
colour and spectral velocities \citep{wan09, fol11,blon12, fol12}. As of yet, none of these spectral trends
have significantly out-performed the standard photometric indices used
in constructing Hubble diagrams. Trends between feature velocity and host galaxy stellar mass have also been identified  \citep{fol12} that may hint at a connection between
spectral properties and the underlying progenitor systems.
\cite{ste11} recently provided evidence of systematically blue-shifted absorption components in SNe Ia, likely related to pre-explosion outflow, favouring the single-degenerate progenitor scenario in spiral galaxies. This work has been extended by \cite{fol12c} to show that the SNe with blue-shifted circumstellar absorption features, on average, display higher ejecta velocities  and redder maximum light colours than normal SNe Ia.

The above results and correlations are very important,
both in our understanding of the physical diversity of the SN Ia phenomenon
as well as the potential for biases in the use of SNe\,Ia as cosmological
probes. However, the local UV samples have been, to date,
modest in size. In this paper, we rectify this shortcoming.

Our goals are several. Using a new, unbiased, low-$z$ sample we aim to verify whether there is evolution with redshift in the NUV spectra, as well as investigate the claimed correlations between NUV spectral, light curve and host galaxy properties, and their effect on the Hubble residuals. To do this, we use the maximum light spectra of 32 low
redshift SNe Ia ($0.001<z<0.08$) at wavelengths down to
$\simeq$2900\AA\ (NUV) obtained using the \textit{HST}.  Our new
sample expands the initial study of 12 events presented in
\cite{coo11}, and now also includes additional multi-colour light curve
data, which are used to colour-correct the spectra. All of our NUV
spectra were obtained with the Space Telescope Imaging Spectrograph
(STIS) on \textit{HST}; 28 were obtained during \textit{HST} Cycle 17
(GO 11721, PI: Ellis), and 4 during Cycle 18 (GO 12298, PI: Ellis).
Although data down to $\sim$3200 \AA\ can be obtained from the ground,
these data are notoriously difficult to calibrate near the atmospheric
cut-off. The two key advantages of \textit{HST} are an accurate
relative flux calibration over the entire wavelength range, and
minimal host galaxy contamination of the spectra due to the narrow
slit employed.

We primarily select our SNe Ia from the Palomar Transient Factory
\citep[PTF,][]{rau09, law09}, a low-$z$ rolling search that
selects SNe in a similar manner as those found in high-$z$
surveys, without any bias towards a particular host type. We use these
data to compare to the NUV spectra of 36 intermediate-redshift ($0.4<z<0.9$) SNe
Ia from the Supernova Legacy Survey (E08) to look for
evolution in the SN Ia properties with redshift and compare the
dispersion of their spectra, for the first time using an unbiased low-$z$ sample. This homogenous low-$z$ sample can
also be split into mean spectra based on their light curve and host
galaxy properties.  The positions (velocities) of the NUV spectral
features identified in the individual spectra can also be measured
qualitatively and compared with their stretches, colours, Hubble
residuals and host galaxy properties.

The plan of the paper follows. $\S$~\ref{obs_data} introduces the data
used in this paper, including the \textit{HST} spectra, the
ground-based optical light curves, and the SN Ia light curve fitting
technique. This section also gives information on the host galaxies
and the method for the calculation of Hubble residuals.
$\S$~\ref{results1} details the results of the mean spectrum
comparisons, including those split by redshift, as well as SN and host
galaxy properties. $\S$~\ref{results2} describes the results of
wavelength and velocity measurements of spectral features and how they relate to
observed SN quantities and Hubble residuals.  The results of the
previous two sections will be discussed in $\S$~\ref{discussion},
including the implications for the use of NUV spectral features as
further calibrators of the Hubble diagram, how these correlations
relate to possible progenitor scenarios, and the cause of the observed evolution with redshift and dispersion at
NUV wavelengths.  Throughout this paper, where necessary we assume a
Hubble constant, $H_0=70$\,km\,s$^{-1}$\,Mpc$^{-1}$.

\section{Observations and Data Reduction}
\label{obs_data}

\subsection{Sample selection}
\label{sec:sample-selection}

\begin{figure*}
\includegraphics[width=16cm]{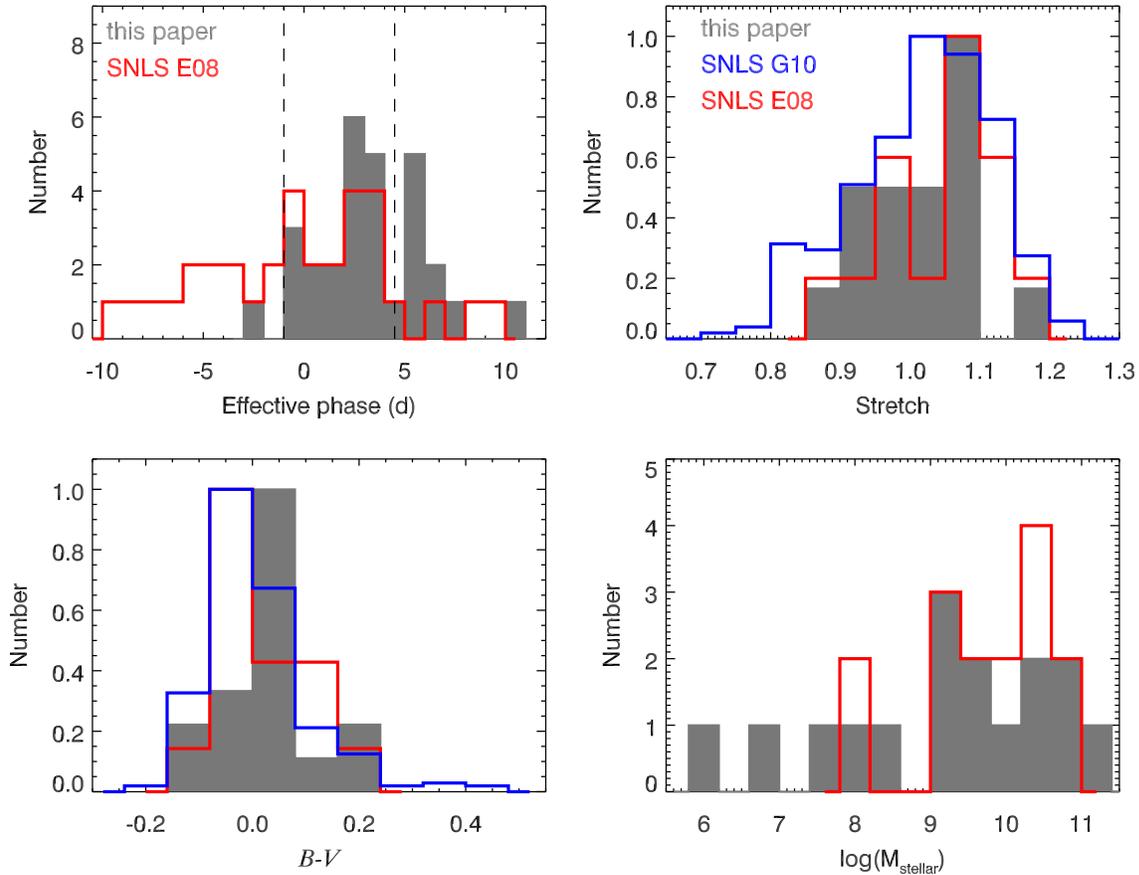} 
\caption{Comparisons of the SN Ia properties for the different samples
  used in this paper. In the top left panel, the effective phase (the SN phase divided by stretch)
  distribution of the sample in this paper (gray), along with the
  sample of \protect \cite{ell08} (E08, red) is shown.  The vertical,
  dashed lines mark the positions of the phase cuts at -1.0 and +4.5 d
  that are used to choose the SNe Ia that enter the mean spectra.  The
  top right and bottom left panels show the stretch
  and $B-V$ colour distributions, respectively, for the phase-selected
  low-$z$ sample (gray), the phase-selected sample from E08 (red) and that of the full SNLS3 SN Ia sample from
  \protect \cite{guy10} (G10, blue). These two plots have had their
  distributions normalised to one for a clearer comparison.  The lower
  right panel shows the host galaxy stellar mass distributions, (M$_{\rm{stellar}}$),
  respectively for the phase-selected low-$z$ sample (gray) and
  the sample of \protect E08 (red), as calculated from fits
  to the host galaxy magnitudes. If the host galaxy was not detected
  (PTF10ufj), a value of M$_{\rm{stellar}}$ of 10$^{6}$ \msun\ was
  assigned.}
\label{dist_dia}
\end{figure*}

The bulk of our NUV spectra of SNe Ia come from a `non-disruptive' Target of
Opportunity (ToO) programme (GO 11721, PI: Ellis). The scheduling
constraints of \textit{HST} meant that the SN Ia targets had to be
submitted for scheduling $\sim$7--16 d before the required
maximum-light spectrum \citep[for further details of our
techniques, see][]{coo11}. This requires very early detection and
spectroscopic classification of potential candiates, which were
discovered by the PTF, and classified by a mixture of scheduled and
ToO spectroscopic time on a large number of optical telescopes,
detailed below.

PTF is a wide area, rolling search optical survey using the Samuel
Oschin 48-in Telescope (P48) located at the Palomar Observatory \citep{rau09, law09}.
Variable cadences in the range of minutes to $\sim$5 d are routinely used to
identify SN candidates very soon after explosion \citep[e.g.][]{gal11}. 27 of the 32 low-$z$
SNe Ia studied here were discovered by PTF, with 8 of the PTF SNe being
identified using the ``Galaxy Zoo: Supernova'' citizen science project
\citep{smi11}. Five other SNe Ia were discovered by other searches: SN
2009le \citep[Chilean Automatic Supernova Search;][]{pig09}, SN 2010ju
\citep[Lick Observatory Supernova Search (LOSS);][]{cen10}, SN 2010kg
\citep[LOSS;][]{nay10}, SN 2011by \citep[Xingming Observation Sky
Survey;][]{jin11} and SN 2011ek \citep[][]{nak11}. Table
\ref{opt_spec} details the optical classification spectra of the
sample, the redshifts of the SNe and their host galaxy stellar masses (M$_{\rm{stellar}}$, which will be discussed further in Section \ref{host_galaxy}).

30 NUV spectra of 30 different SNe were obtained during the main cycle
17 programme, 28 of which were ``normal'' SNe Ia, one of which was the
unusual SN Ia PTF10ops \citep[detailed in][]{mag11}, and one of which
was ultimately classified as a SN Ic (PTF10osn). These latter two
events are not studied in this paper.  During \textit{HST} Cycle 18, a
follow-on \textit{HST} programme (GO 12298, P.I.: Ellis) studied the time evolution in the NUV and far-UV
spectra of a smaller sample of SNe Ia from soon after explosion to after 
maximum light.  This programme is complementary to the Cycle 17
programme and will be presented in detail in further publications. However, we have supplemented our sample of the 28 ``normal'' cycle 17
SNe Ia maximum light spectra with four maximum light spectra from this
programme: PTF10ygu (Hachinger et al.~in prep.), PTF11kly/SN 2011fe \citep{nug11}, SN 2011ek and
SN 2011by. This brings our sample to a total of 32 maximum-light NUV
spectra, by far the largest NUV spectral sample of low-$z$ SNe Ia to
date.

The light curve derived properties (date of maximum, $B$-band magnitude at maximum, stretch and $B-V$ colour at maximum) of the sample are described in Table \ref{uv_spec_tab}. We
will compare our sample to that of E08, which contains a
large sample of intermediate-$z$ SN Ia UV spectra that were obtained
using Keck+LRIS. Their sample of SNe Ia was discovered by the
Supernova Legacy Survey \citep[SNLS;][]{guy10} and are in the redshift
range of $\sim$0.4--0.9, with a mean redshift of 0.6. We extend the intermediate-$z$ sample of
E08 to include nine further SNLS SNe Ia spectra obtained with
the same Keck setup (although not part of the E08 sample) and
reduced using the same method as described in E08. Throughout
this paper, references to the intermediate-$z$ E08 sample
will include these nine additional SN spectra. A comparison of the effective phase (the phase of the NUV spectrum divided by the SN stretch), stretch, $B-V$ colour at maximum, and M$_{\rm{stellar}}$ distributions of the low-$z$ sample to that of the intermediate sample of E08 is shown in Fig.~\ref{dist_dia}, along with the full three-year SNLS
SN Ia sample \citep[SNLS3;][]{guy10,con11,sul11}.

Using a Kolmogorov-Smirnov (K-S) test, we find that there is a very low probability of our low-$z$ \textit{HST} sample and the intermediate-$z$ sample of E08 having their effective phase, stretch, $B-V$ colour at maximum, and M$_{\rm{stellar}}$ distributions drawn from different parent populations (using a phase range of -1.0--4.5 d, the choice of which will be discussed in Section \ref{selection_mean}). This suggests that the chosen low- and intermediate-$z$ samples are well-matched and any evolution in their spectral properties should not be due to differences in the phase and light curve properties of the samples being studied. Luminosity biases that could occur because of brighter SNe Ia being discovered preferentially at intermediate-$z$ would be manifested as a stretch or colour bias, which is not seen.

\begin{table*}
 \caption{Log of optical classification and \textit{HST} spectra, along with the SN redshifts and host galaxy log(M$_{\rm stellar}$) values.}
 \begin{tabular}{@{}lccccccccccccccccccccccccccccccc}
  \hline
  \hline
SN name& R.A. &Decl. & Class. Telescope/&MJD &MJD \textit{HST} &CMB redshift ($z$)$^c$& Log(M$_{\rm{\rm{stellar}}}$)\\ 
&(J2000)&(J2000)&CBET&Classification&spectrum&&(\msun)&\\ 
\hline
    PTF09dlc  &21:46:30.10& +06:25:09.2 &P200+DBSP	&55063.5 & 55076.7	&0.0666$\pm$0.0005	&           9.01 \\
    PTF09dnl  &17:23:41.79 &+30:29:49.6	&P200+DBSP& 55063.5& 55076.1	&0.019$\pm$0.001		&   7.84\\
    PTF09dnp     & 15:19:24.37& +49:29:56.3&P200+DBSP&55063.2& 55076.6	&0.037559$\pm$0.00001$^1$	& 11.10\\
    PTF09fox   	&23:20:47.98 & +32:30:08.2& Keck+DEIMOS&55123&55134.1		&0.0707$\pm$0.0005	&    10.39\\
    PTF09foz   	&00:42:11.73& -09:52:52.7& P200+DBSP&55126&55134.2		&0.05331$\pm$0.00002	& 10.94\\
    PTF10bjs   	&13:01:11.23& +53:48:57.3&WHT+ISIS&55250.2& 55263.7		&0.0303$\pm$0.0003	&    6.85\\
    PTF10fps$^b$   	&13:29:25.06& +11:47:46.5&P200+DBSP&55303& 55319.5		&0.022513$\pm$0.000009$^1$& 10.90\\
    PTF10hdv 	&12:07:45.37 &+41:29:27.9&KECK+LRIS&55331&   55347.6	&0.0542$\pm$0.0003	& 7.50\\
    PTF10hmv$^b$	&  12:11:32.99& +47:16:29.8&P200+DBSP&55337&55354.3	&0.033$\pm$0.001		& 8.34\\
    PTF10icb   	& 12:54:49.21 &+58:52:54.7&Gemini N+GMOS&55349&55361.4		&0.0088$\pm$0.0003	&   9.66\\
    PTF10mwb$^b$	&17:17:49.97& +40:52:52.1 &Gemini N+GMOS&55367& 55390.4	&0.0312$\pm$0.0005	&   9.36\\
    PTF10ndc$^a$      &17:19:50.18 &+28:41:57.5&P200+DBSP&55384& 55396.5	&0.0817$\pm$0.0005	& 9.05\\
    PTF10nlg &16:50:34.48& +60:16:35.0&Keck+LRIS&55384.4& 55397.3&0.056$\pm$0.005		&10.11\\
    PTF10pdf   &13:24:30.98 &+42:05:20.3&P200+DBSP&55396&55410.1	&0.0764$\pm$0.0005 	&  11.08\\
    PTF10qjl            &16:39:59.33 &+12:06:25.5&WHT+ISIS&55411.0 & 55424.4 	&0.0579$\pm$0.0005	&    9.10\\
    PTF10qjq           &17:07:12.39 &+35:30:35.3	&WHT+ISIS&55411.0&  55424.4 	&0.0288$\pm$0.0005	&   10.14\\
    PTF10qyx$^b$        &  02:27:12.06& -04:32:04.8	&Lick+KAST&55419.5&  55431.8 	&0.065$\pm$0.005$^2$&   6.00\\
    PTF10tce$^{a,b}$    &	23:19:10.36& +09:11:54.2&Gemini N+GMOS&	 55433&55445.7 	&0.039716$\pm$0.000017$^3$&   10.60\\
    PTF10ufj$^{a,b}$     	&02:25:39.13 &+24:45:53.2&P200+DBSP&55444&55459.1		&0.076$\pm$0.005$^2$	&   6.00\\
    PTF10wnm 	&00:22:03.61& +27:02:26.2&VLT+XSHOOTER&55467& 55480.6		&0.0645$\pm$0.0001	&	   10.52\\
    PTF10wof$^a$   	&23:32:41.84& +15:21:31.7&Gemini N+GMOS&55468& 55480.1		&0.0514$\pm$0.0005	&   10.18\\
    PTF10xyt$^a$    &23:19:02.42 &+13:47:26.8&KPNO+RC Spec &55480.5 &55494.3 	&0.0484$\pm$0.0003	&   9.75\\
    PTF10ygu$^{a,b}$&09:37:30.29& +23:09:33.0  &Gemini N+GMOS&55482& 55495.4	&0.0260$\pm$0.0005	&   11.20\\
    PTF10yux$^a$      &23:24:13.39& +07:13:42.7&Gemini N+GMOS& 55487   &   55502.3	&0.0559$\pm$0.0005	&   11.22\\
    PTF10zdk$^a$         & 02:14:07.33 &+23:37:50.2 &WHT+ACAM&55502.0	&55515.1		&0.032$\pm$0.001		&    	9.70\\
    PTF10acdh     &09:43:07.58& +09:39:31.3  &P200+DBSP&55543&55558.1		 &0.0606$\pm$0.0002$^1$	&  10.82\\
    PTF11kly           &14:03:05.83 &+54:16:25.2&LT+FRODOSPEC  &55797.9&55814.4  	&0.001208$\pm$0.000005&  --\\
    SN2009le &02:09:17.14&-23:24:44.8 &CBET 2022/2025 &55151&55165.2&0.01703$\pm$0.000030$^3$ &   11.04\\
    SN2010ju&05:41:55.99 & +18:29:51.0&CBET 2549/2550&55516.3&55529.6&0.015348$\pm$0.000013$^3$ &    10.93\\
    SN2010kg&	 04:40:08.40 & +07:21:00.0 &CBET 2561&55531&55543.3		 &0.013968$\pm$0.000013$^3$ & --\\
    SN2011by&11:55:45.56 & +55:19:33.8 &CBET 2708&55678.5&55690.3	  &0.003402$\pm$0.000005$^3$ &   9.27\\
    SN2011ek&02:25:48.89&+18:32:00.0 &CBET 2783&55780.2&55792.4&0.004206$\pm$0.000010$^3$ &   --\\
\hline	
 \end{tabular}
  \label{opt_spec}
 \begin{flushleft}
$^a$`Galaxy Zoo Supernovae project' discovered SNe.\\
$^b$IAU designated name, PTF10fps=SN 2010cr, PTF10hmv=SN 2010dm, PTF10mwb=SN 2010gn, PTF10qyx=SN 2010gy=PS1-100385, PTF10tce=SN 2010ho, PTF10ufj=SN 2010hs, PTF10ygu=SN 2010jn\\
$^c$If not otherwise marked, $z$ values are measured from host galaxy features. $^1$SDSS, $^2$template fit, $^3$NED\\
P200+DBSP: Palomar 200-inch telescope with the Double Spectrograph \protect \citep{oke82}.\\
Keck+DEIMOS: Keck 10-m with the Deep Imaging Multi-Object Spectrograph \citep{fab03}.\\
WHT+ISIS: 4.2-m William Herschel Telescope with the Intermediate dispersion Spectrograph and Imaging System.\\
Keck+LRIS: Keck 10-m with the Low-Resolution Imaging Spectrometer \citep{oke95}.\\
Gemini N+GMOS: Gemini North with the Gemini Multi-Object Spectrograph \citep{hoo04}.\\
Lick+KAST: Lick Observatory Shane 3-m telescope with the KAST double spectrograph.\\
VLT+XSHOOTER: Very Large Telescope with the XSHOOTER spectrograph \citep{dod06,ver11}.\\
KPNO+RC Spec: KPNO 4-m Mayall telescope with the RC spectrograph.\\
WHT+ACAM: 4.2-m William Herschel Telescope with the Auxiliary-Port Camera.\\
LT+FRODOSPEC: Liverpool Telescope with the Fibre-fed Robotic Dual-beam Optical Spectrograph.\\
\end{flushleft}
\end{table*}

\begin{table*}
 \caption{Log of photometric properties of the SN Ia sample.}
 \begin{tabular}{@{}lccccccccccccccccccccccccccccccc}
  \hline
  \hline
SN name&MJD& Effective&$B$-band&Stretch &$B-V$& \\ 
& $B$ max&phase (d)$^b$& max (mag)&&\\ 
\hline
PTF09dlc    &   55073.7$\pm$0.2    &   2.8    & 18.04$\pm$0.07   &    1.05$\pm$0.03   &$-0.16\pm$0.05     	\\
PTF09dnl    &   55074.8$\pm$0.1  &     1.3     & 15.80$\pm$0.02 &      1.05$\pm$0.02   &$-0.02\pm$0.01	\\
PTF09dnp  &     55071.0 $\pm$0.2 &      5.8 &     16.60$\pm$0.07&       0.96$\pm$0.05  &  $-0.15\pm$0.06	\\
PTF09fox    &   55131.7$\pm$0.2     &  2.6    & 18.25$\pm$0.06 &     0.92$\pm$0.04  & 0.00$\pm$0.04	\\
PTF09foz    &   55131.8$\pm$0.3   &    2.8    & 17.81$\pm$0.10 &     0.87$\pm$0.06   & 0.03$\pm$0.08	\\
PTF10bjs    &   55261.6$\pm$0.2   &    1.9    & 16.01$\pm$0.03 &      1.08$\pm$0.02   & $-0.09\pm$0.02	\\
PTF10fps    &   55311.7$\pm$0.3   & 10.6   & 16.66$\pm$0.05   &   0.73$\pm$0.03     &0.14$\pm$0.04				\\
PTF10hdv   &    55344.1$\pm$0.3   &    3.3    &  17.56$\pm$0.03    &   1.05$\pm$0.07&    0.09$\pm$0.03	\\
PTF10hmv  &     55351.4$\pm$0.1  &     2.5  &   17.33$\pm$0.06    &   1.15$\pm$0.01    & 0.19$\pm$0.04	\\
PTF10icb    &   55360.6$\pm$0.1    &   0.8     &   14.48$\pm$0.03   &   0.99$\pm$0.03   & 0.06$\pm$0.02	\\
PTF10mwb &      55390.7$\pm$ 0.1 &     $-$0.4&      16.81$\pm$0.04 &     0.94$\pm$0.03  &   0.03$\pm$0.03	\\
PTF10ndc   &    55390.3$\pm$0.1   &   5.8   &  18.41$\pm$0.02&       1.05$\pm$0.03  & $-0.05\pm$0.02	\\
PTF10nlg    &   55391.5 $\pm$0.2    &   6.2   &  18.60$\pm$0.03&      0.94$\pm$0.05    & 0.13$\pm$0.03  	\\
PTF10pdf$^a$&  55407.4$\pm$0.3	 &  2.2	  &  --	& 1.23$\pm$0.03&-- 	\\
PTF10qjl     &  55418.9 $\pm$0.1     &  5.9    & 17.79$\pm$0.03 &     0.93$\pm$0.02  &  $-0.10\pm$0.02	\\
PTF10qjq    &   55421.0$\pm$0.1     &  3.5     & 16.08$\pm$0.02 &     0.96$\pm$0.02  &  0.08$\pm$0.02	\\
PTF10qyx    &   55426.1$\pm$0.1    &   6.8  &   18.21$\pm$0.02 &     0.85$\pm$0.01    &$-0.10\pm$0.02	\\
PTF10tce     &  55442.0$\pm$0.1     &  3.5    &  17.15$\pm$0.03 &      1.07$\pm$0.02  &  0.03$\pm$0.02	\\
PTF10ufj      & 55456.5$\pm$0.2      & 2.7    &  18.31$\pm$0.09  &    0.95$\pm$0.02   & $-0.08\pm$0.06     	\\
PTF10wnm  &     55476.5$\pm$0.1 &  4.1    & 18.17$\pm$0.02   &    1.01$\pm$0.03   & 0.04$\pm$0.02	\\
PTF10wof    &   55474.2$\pm$0.1   &    5.9  &   17.93$\pm$0.07 &     0.99$\pm$0.04  &  0.10$\pm$0.04	\\
PTF10xyt     &  55490.9$\pm$0.2     &  3.2   &  18.40$\pm$0.04   &    1.07$\pm$0.04   &  0.19$\pm$0.03	\\
PTF10ygu    &   55495.8$\pm$0.1   &   $-$0.3 &    17.29$\pm$0.04 &      1.07$\pm$0.02  &   0.44$\pm$0.03	\\
PTF10yux    &   55496.4$\pm$0.1   &    7.1   &   18.44$\pm$0.05 &     0.83$\pm$0.01   &  0.16$\pm$0.02	\\
PTF10zdk$^a$	   &			--		&	--	  & 		--		& 	--&--			\\
PTF10acdh$^a$ &     55550.1$\pm$0.1	&   9.1 & 		--		& 	0.80$\pm$0.02   & --	\\
PTF11kly     &  55814.9$\pm$0.04     &  2.8     &    10.12$\pm$0.01&       1.00$\pm$0.01  &  0.06$\pm$0.01	\\
SN2009le  &    55164.9$\pm$0.5   &    0.3   &  15.15$\pm$0.06 &     1.08$\pm$0.01   & 0.06$\pm$0.07	\\
SN2010ju  &    55523.9$\pm$0.4  &     5.5   &  15.79$\pm$0.23 &     1.05$\pm$0.58   & 0.23$\pm$0.06	\\
SN2010kg &               		--		&	--	  & 		--		& 	--&--			\\
SN2011by &     55690.6$\pm$0.1 &     $-$0.3  &   12.97$\pm$0.07&      0.93$\pm$0.02    & 0.02$\pm$0.01	\\
SN2011ek &     55788.9$\pm$0.1  &     3.7   &   13.84$\pm$0.15&     0.90$\pm$0.02    & 0.18$\pm$0.04	\\
\hline	
 \end{tabular}
  \label{uv_spec_tab}
 \begin{flushleft}
$^a$No multi-colour light curves were obtained for PTF10pdf, PTF10zdk, PTF10acdh and SN2010kg.\\
$^b$ Effective phase is the measured phase divided by the stretch.\\
\end{flushleft}
\end{table*}

\subsection{UV spectroscopy}
\label{uv_spec}

\begin{figure*}
\includegraphics[width=17cm]{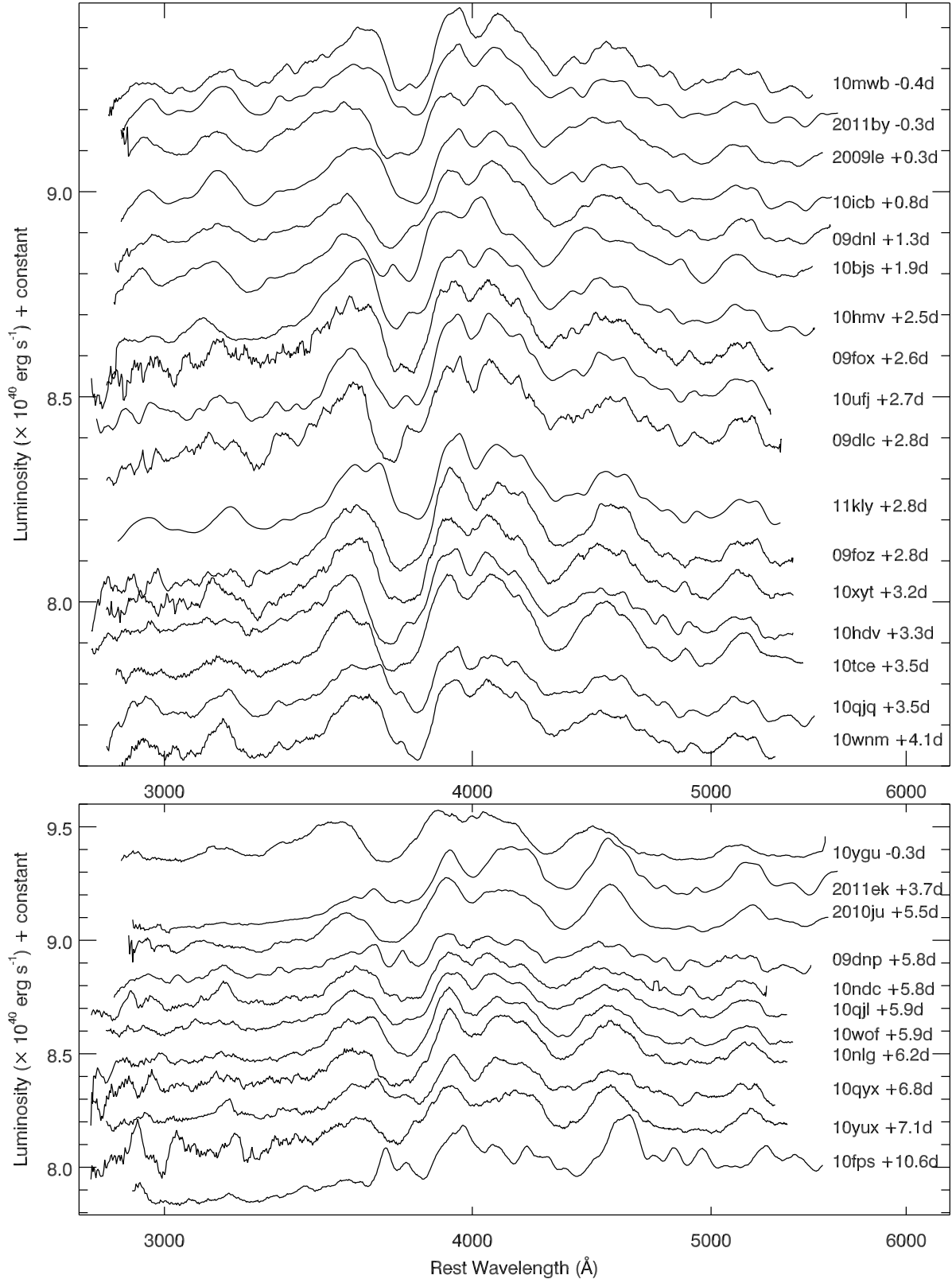}
\caption{Our \textit{HST} NUV spectral sample. The top panel shows the SN Ia spectra entering the mean spectrum comparison, while the bottom panel shows the SNe that have been removed due to $B-V$ colour cuts (PTF10ygu, SN 2011ek), along with those removed due to the effective phase cuts (-1.0 to +4.5 d). For each event, the effective phase is listed next to the SN name. The four SNe Ia in the sample for which $B-V$ colours could not be calculated are excluded from the plot since colour corrections could not be applied.}
\label{uv_spec_fig}
\end{figure*}

The \textit{HST} spectra were obtained using STIS and the G430L grism, giving a wavelength
coverage of 2900--5700\AA\ with a dispersion of 2.73 \AA\ per pixel and
a plate scale of 0.051\arcsec\ per pixel. 
Fig.~\ref{uv_spec_fig} shows the NUV spectra which are discussed in
this paper. All the spectra are publicly released via the WISeREP portal\footnote{http://www.weizmann.ac.il/astrophysics/wiserep/;} \citep{yar12}.

The spectra were reduced in a manner broadly following that of \cite{coo11}.
The spectra were downloaded from the \textit{HST} archive using the
on-the-fly reprocessing (OTFR) pipeline. This gives fully calibrated
1D spectra (`sx1' files), where the reduction and extraction is
optimised for point sources.
The OTFR pipeline uses the latest calibration files and data
parameters to perform initial 2D image reduction such as image
trimming, bias and dark current subtraction, cosmic-ray rejection via
\textsc{CRSPLIT} and flat-fielding. It then performs wavelength and
flux calibrations, including correcting the flux for imperfect charge
transfer efficiencies along the chip. We applied further cosmic ray
removal using LACOSMIC \citep{van01} and corrected the spectra for
Milky Way extinction using an $R_V$ value of 3.1, the dust maps of
\citet{sch98}, and the Milky Way dust extinction relation of \citet[][CCM]{car89}.  Most of the
SNe Ia in the sample had low Milky Way extinction values of $E(\bv)$ $<0.2$,
although two, SN 2011ek and SN 2010ju had values of 0.35 and 0.42
respectively.  SN 2011ek had an unusually red UV minus optical colour
and the \textit{HST} spectrum of SN 2010ju is outside the phase range discussed in the rest of the paper. Both have been excluded from further analysis.

We assess the possibility of host galaxy contamination in our \textit{HST}/STIS
spectra using the acquisition images taken with each spectrum. These use the
STIS/CCD long-pass filter and have exposure times ranging from 5--35 s,
depending on the apparent magnitude of the SN. In all cases, the SN is
by far the brightest object in the image, and only 7 of the images
show any visual trace of the host galaxy. For each SN, we estimate the
possible contamination by placing circular apertures of 0.2\arcsec\
diameter (the slit width, approximately 4 STIS/CCD pixels) at the
position of the SN, and also along the circumference of a circle of
radius 0.6\arcsec\ centered on the SN position. These latter apertures
measure the host galaxy flux near to the SN position but are outside
the extent of the STIS/CCD point spread function (PSF). We take the brightest of these host
galaxy apertures as conservatively representing the flux attributable
to the host galaxy. We then calculate the percentage excess in the
brightness of the SN over its host. For 30 SNe (94 per cent of the sample), the SN light through the slit is more than 100 times the estimated brightness of the host through the slit.  Even for the worst case SN (PTF10xyt), the SN flux through the slit is $\sim$30 times that of the host galaxy flux. Therefore, we consider the host galaxy contamination to be negligible in our analysis.

Spectral line identifications were made for all the \textit{HST} spectra in the sample using the SN spectral fitting code, \textsc{synapps}, \citep{tho11}. This code is based on \textsc{synow} \citep{fis00} and has the same assumptions, and so is limited to identification of features and estimates of their ejection velocities. It does not provide measurements of ion species abundances. The main advantage of \textsc{synapps} over {\sc synow} is that it optimises automatically over the input parameters to perform a highly parameterised fit to the observed spectra. The outputs of \textsc{synapps} include ions contributing to the various spectral features and the velocities of the individual ions used in the fit.

\begin{figure*}
\includegraphics[width=18cm]{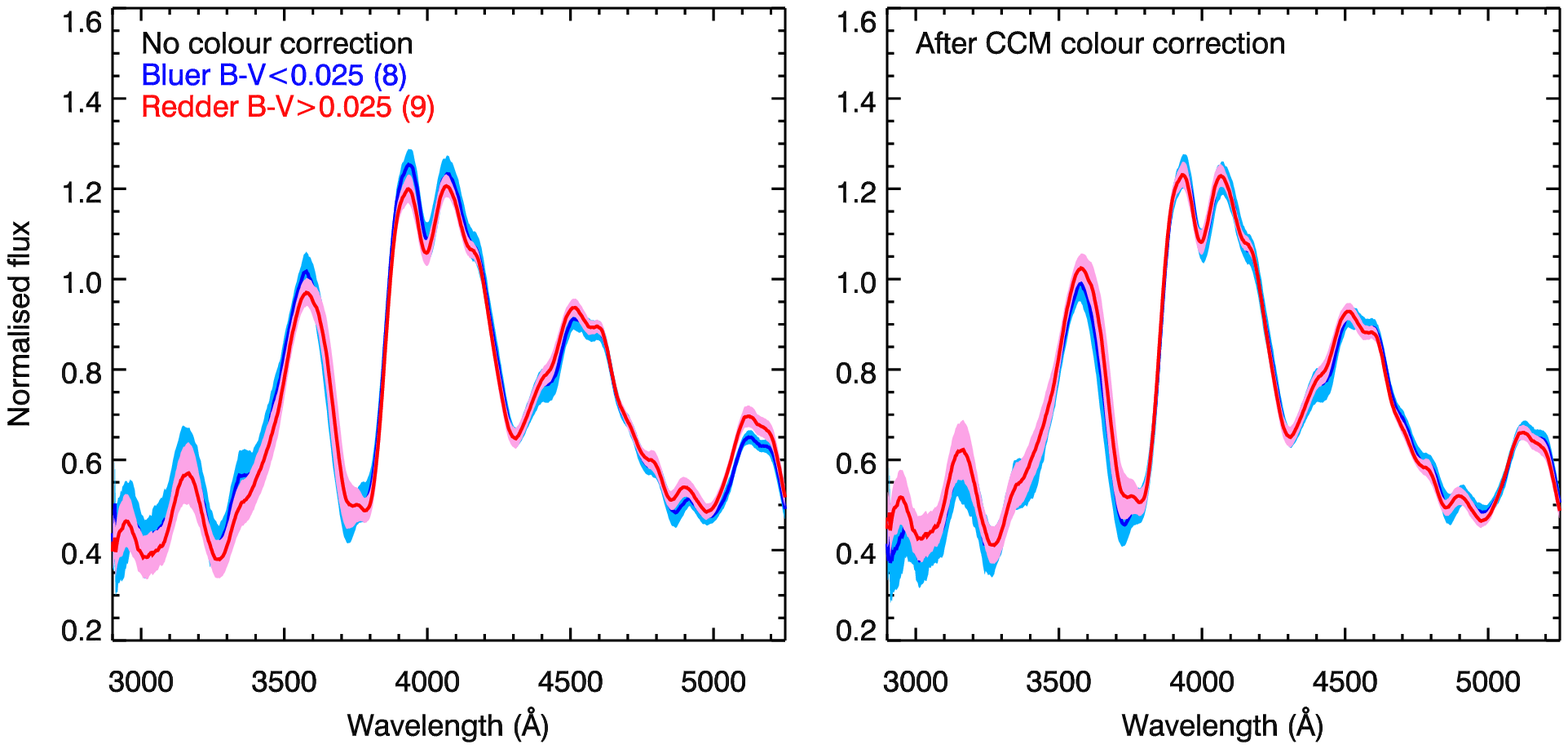}
\caption{The mean spectra of the \textit{HST} sample are shown for blue and red SNe Ia as the solid lines in these plots. The left panel shows the SNe before colour-correction, when split based on the $B-V$ colour obtained from light curve fits, with blue ($B-V<0.025$) and red ($B-V>0.025$). In the right panel, the CCM-colour corrected SN mean spectra are shown. The mean spectra were constructed as described in Section \ref{selection_mean}. The shaded regions represent the 90 per cent confidence levels from a bootstrap resampling.}
\label{nocc}
\end{figure*}

\subsection{Optical photometry}
\label{opt_phot}

Multi-colour light curves were obtained at the robotic 2-m Liverpool
Telescope \citep[LT;][]{ste04} located at the Roque de Los Muchachos
Observatory on La Palma. The optical imager, RATCAM was used with $gri$ filters, similar to those used in the Sloan Digital Sky Survey \cite[SDSS;][]{yor00}. RATCAM has a field of view of 4.6$'$ $\times$ 4.6$'$ with a pixel scale of 0.135$''$ (unbinned).

Deep references images of the SN fields were obtained at $>$310 d post explosion at the LT for all the SNe listed in Table \ref{uv_spec_tab} on photometric nights. These were subtracted from the SN images at each epoch to remove host galaxy contamination. This subtraction involves registering and combining five reference images with individual exposure times of 150 s to obtain a stack. For each SN image, the reference image was flux-scaled to it and then subtracted off using a PSF-matching routine. Where possible, the zero-points were calculated directly using SDSS stars in the fields of the SNe to calibrate to the SDSS photometric system. This was not possible for PTF10nlg and SN2009le, which were not in the SDSS footprint. These were instead calibrated using standard star fields, obtained on photometric nights, observed before and after the deep SN reference images. This is a preliminary calibration, and the final low-$z$ sample of SNe that PTF will obtain will all be fully calibrated to SDSS Stripe 82 fields. 

$R$-band light curves for the SNe Ia sample were also obtained using the PTF search telescope, the P48. Pre-explosion reference images were available for the SNe discovered with PTF. The P48 data were reduced by the Infrared Processing and Analysis Center (IPAC)\footnote{http://www.ipac.caltech.edu/} pipeline (Laher et al.~in prep.), and the light curves measured using a custom-built pipeline, similar to the LT pipeline detailed above. The photometric calibration was performed to SDSS following a similar, but independent, method to \cite{ofe12}.

Due to bad weather and scheduling difficulties, suitable LT+RATCAM light curves were not obtained for PTF10pdf, PTF10acdh, PTF10zdk and SN 2010kg. For some SNe, it was possible to supplement the light curves with data from the Faulkes Telescope North (FTN) on Haleakala, of the Las Cumbres Observatory Global Telescope\footnote{http://lcogt.net/} (LCOGT). These data have been reduced and calibrated in a similar manner to that described above for the LT data. Multi-colour light curves of PTF11kly (SN 2011fe) were obtained with the Byrne Observatory at Sedgwick Reserve (BOS) 0.8 m Telescope in $g'r'i'$ band filters and calibrated to the SDSS system.

\subsection{Light curve fitting}
\label{sec:light-curve-fitting}

The optical $gRri$-band light curves were analysed using the SiFTO
light curve fitting code \citep{con08}, which produces values for the stretch, maximum $B$-band magnitude, $B-V$ colour at maximum, and time of maximum light for each SN.
These photometric parameters are listed in Table \ref{uv_spec_tab}.  SiFTO
uses a time-series of spectral templates that are adjusted to recreate
the observed colours of the SN photometry at each epoch, while also
adjusting for Galactic extinction and redshift (i.e., the
$k$-correction). The P48 $R$-band data is not used in the $B-V$ colour measurements, just for constraining the date of maximum and the stretch. Provided there are measured colours for a SN, SiFTO can
also interpolate to obtain the peak magnitude in a chosen rest-frame
filter. The magnitudes and colours quoted throughout the paper have been measured at $B$-band maximum. The colour output of SiFTO is $C$, a weighted average of the $U-B$ and $B-V$ colours at $B$-band maximum. However, because our bluest filter is the $g$-band, SiFTO $C$ and $B-V$ are equivalent for our dataset. Therefore, we will use the term $B-V$ throughout to mean the $C$ colour at $B$-band maximum light. Tests have been performed in \cite{guy10} comparing SiFTO
and the popular SALT2 \citep{guy07} light curve fitter, and good agreement is
found between their outputs, despite the differences in their methods.
The effective phase values quoted in Table \ref{uv_spec_tab} are the
measured phases divided by the stretch as measured from the light
curve fits. A comparison analysis has been performed using the phases before correcting for stretch and does not change the results.

\subsection{Colour correction of UV spectra}
\label{colour_cor}

Dust extinction towards SNe Ia can affect the measured `colour',
with different amounts of extinction resulting in varying colours
being observed. If the intrinsic colours of SNe Ia were assumed to be
the same for all objects and Milky Way-like dust was responsible for
the observed variations in $B-V$ colour, then when using the Milky Way dust
extinction CCM law \citep{car89}, a wavelength parameterisation with a
selective-to-total extinction value of R$_B = 4.1$ (R$_V = 3.1$)  should be applicable.
However, it has been found that when the value of R$_B$ is allowed to
vary, an R$_B \simeq 3$ is favoured by SNe Ia \citep{tri98, tri99,
  con07}. This suggests that either the dust in SNe Ia host galaxies is not consistent with Milky Way-like dust, or that there are intrinsic variations in the colours of SNe Ia, similar to
the observed variation in light curve stretch. Here we correct for the observed $B-V$ colour variations, to remove
the differences in $B-V$ colour (dust and intrinsic colour) that could
affect the measured flux in the spectra.

We investigated two ways to apply the colour correction, using either the SALT \citep{guy07} or the Milky Way dust CCM colour law. The SALT and CCM colour laws have similar forms at optical wavelengths, but are different in the NUV, where the SALT (SALT1, SALT2) colour laws have steeper slopes. Both the SALT colour laws were applied using the SiFTO \textit{C}, corrected to a SALT colour, using the conversion relation from \cite{guy10}.   The chosen value of total-to-selective extinction does not matter since the spectra are normalised in the region 4000--4500 \AA\ and measurements of wavelength positions are also not affected by this value.

 Fig.~\ref{nocc} shows the low-$z$ sample split into two bins at the position of the mean \textit{C} of the sample, before (left panel) and after (right panel) a CCM colour-correction was applied. Before correction, SNe with bluer \textit{C} have bluer spectra, as is expected. After CCM-correction, there is better flux agreement between the \textit{C} split samples. The CCM colour law correction gave a more consistent agreement between spectra than either the SALT1 or SALT2 colour laws, which overcorrected the colour of the NUV spectra making the redder SNe significantly bluer than the originally bluer SNe. E08 used the SALT1 law because of better post-correction agreement between the low and high \textit{C} SNe, but we note that in fig. 8 of E08 the difference between the CCM and SALT1 colour law corrections is small. Therefore,  for consistency the CCM colour law has been used to correct both the low-$z$ and intermediate-$z$ samples.

\begin{figure*}
\includegraphics[width=18cm]{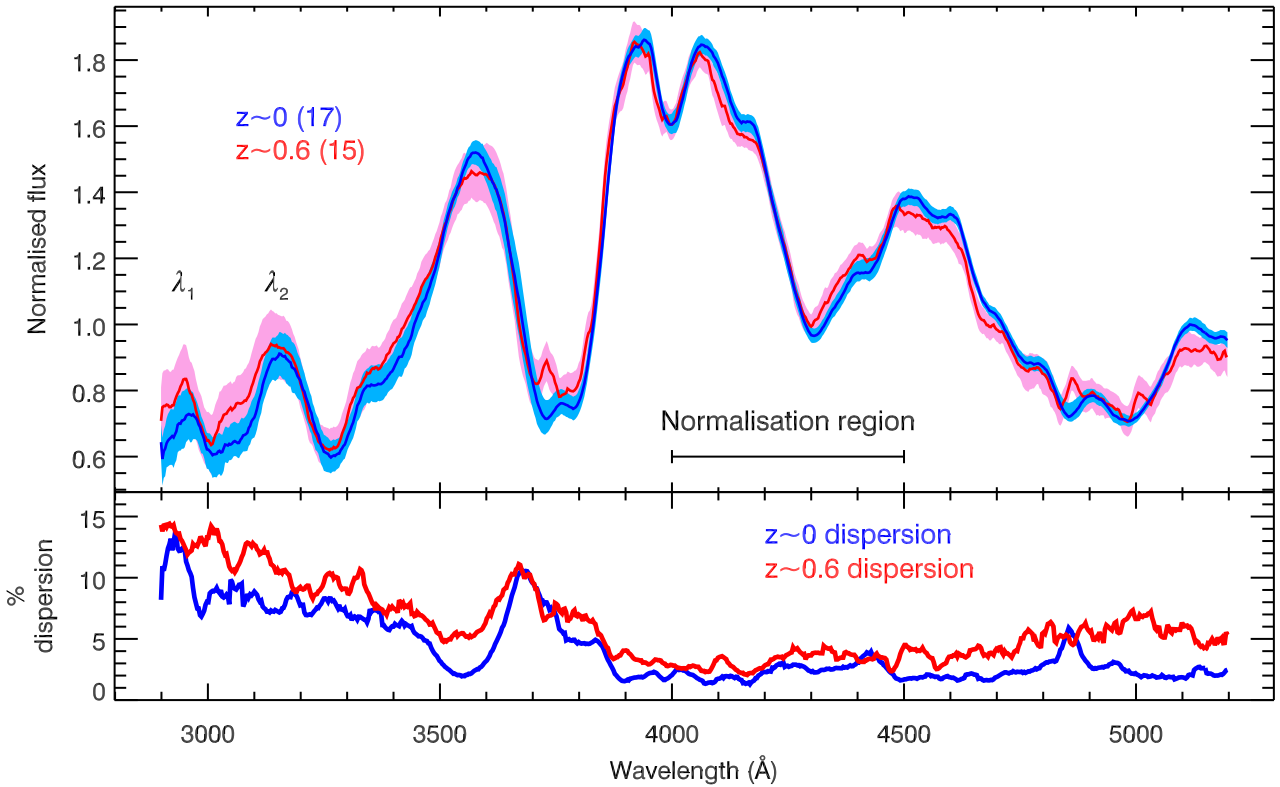}
\caption{Comparison of the mean spectra of the low-$z$ sample (17 SNe Ia) to that of the intermediate-$z$ sample (15 SNe Ia) in the top panel. The shaded regions represent the 90 per cent confidence levels from a bootstrap resampling. The region through which the normalisation was applied is marked, along with positions of the $\lambda_1$ and $\lambda_2$ NUV features.The lower panel shows the percentage dispersion at each wavelength (bootstrap resampling error as a percentage of the flux of the mean) for the \textit{HST} and Keck samples respectively. An increase in the dispersion in the NUV compared to optical wavelengths is seen. A peaked appearance is seen in the dispersion of both samples around $\sim$3700 \AA, which is found to be caused by an increased dispersion in the blue wing of the \Caii\ H\&K feature.}
\label{mean_highlow}
\end{figure*}

\subsection{Host galaxies}
\label{host_galaxy}

The redshift of the host galaxies were obtained in order of preference i) from host galaxy lines identified in the optical SN spectra, ii) from galaxy spectra obtained from the NASA/IPAC Extragalactic Database\footnote{http://nedwww.ipac.caltech.edu/} (NED) or iii) from the SDSS  (labelled `NED' or `SDSS'  in Table \ref{opt_spec}). However, it was not possible to obtain a host galaxy redshift for two SNe, PTF10ufj and PTF10qyx (labelled `template' fit in Table \ref{opt_spec}). A faint host at the position of PTF10ufj was detected in a 200 s $R$-band image obtained with Keck+LRIS, but a 1800 s combined spectrum, obtained with KECK+LRIS using the 600/4000 blue grism and 400/8500 red grating, did not reveal any identifiable features that could be attributed to a host galaxy. For PTF10qyx, no host was detected in a 200 s $R$-band image obtained at KECK+LRIS to a limiting magnitude of $\sim$25.5 mag. 

For these two SNe, redshifts  were obtained using the template matching routine \textsc{superfit} \citep{how05} by selecting the average redshift obtained from all the available spectra for these SNe. This redshift was then used to correct the spectra to the rest-frame. These SNe were excluded from the wavelength shift and velocity analysis in Section \ref{results2} because of the larger uncertainties in their redshifts. 

The host galaxy M$_{\rm{stellar}}$ of our SN Ia sample were calculated using the PEGASE code \citep{leb02}, with SDSS multi-band photometry and the method of \cite{sul10} to fit the SED of the host photometry to a series of galaxy templates. The values of these fits are listed in Table \ref{dist_dia} and the distribution of the host M$_{\rm{stellar}}$ is shown in Fig.~\ref{dist_dia}. For SNe in the sample that do not have SDSS photometry available, the magnitude of the galaxy was estimated using the $gri$-band deep reference images obtained with LT, from which the M$_{\rm{stellar}}$ could be estimated.  

If the host galaxy was not detected (PTF10qyx, PTF10ufj), a M$_{\rm{stellar}}$ of 10$^{6}$ \msun\ was assigned. However, there are three galaxies at separations of $1-5\arcmin$ from the position of PTF10qyx, that may be potential host galaxies. The nearest potential host galaxy, VVDS 020218384 is at a separation of 0.9\arcsec and has a redshift of 0.068, within the uncertainties of the SN redshift value of 0.065$\pm$0.005 \citep{lef05}. This suggests that the host of PTF10qyx may not be very faint/low mass but more massive and distant. However, PTF10qyx are not included in the wavelength shift and velocity studies, and so will not affect our conclusions.

\begin{figure*}
\includegraphics[width=18cm]{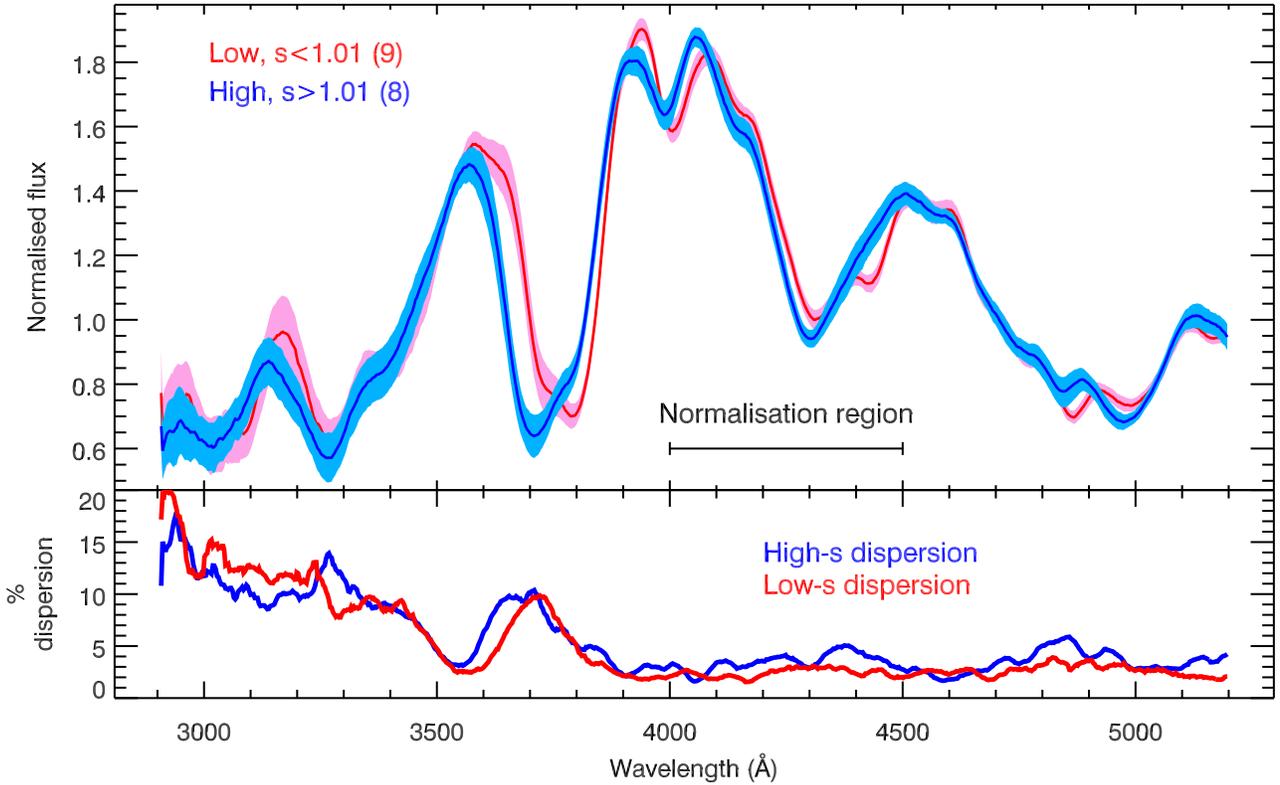}
\caption{The SN Ia low-$z$ subsample split based on stretch, with a cutoff between high (blue) and low (red) stretch of 1.01. The bottom panel shows the dispersion of the high- and low-stretch means. The numbers in parentheses are the number of SN Ia spectra that went into each mean spectrum.}
\label{mean_stretch2}
\end{figure*}

\subsection{Hubble residuals}
\label{make_HR}

For cosmology using SNe Ia, the peak $B$-band magnitude is `corrected' using empirical relation between light curve stretch, light curve colour, and host M$_{\rm{stellar}}$. The residuals on the Hubble diagram after correction for some or all of these parameters are studied to look for possible correlations with spectral features. These could be used to further reduce the scatter in the Hubble diagram or provide alternative correction methods.

The residuals, $m_{B}^{resid}$ are calculated using
\begin{equation}
m_{B}^{resid}=m_{B}^{max}-M_B - 5~log_{10}~D_L - 25  + \alpha (s-1) - \beta C
\label{eqn1}
\end{equation}
where $m_B^{max}$ is the measured B-band magnitude, $D_L$ is the luminosity distance in units of Mpc, $s$ is the stretch and $C$ is the colour of the SN. $\alpha$, $\beta$ and $M_B$ are the coefficients of the best-fit cosmological parameters to the data following the technique in \cite{sul11}. We assume a standard flat-$\Lambda$CDM cosmology ($\Omega_m=0.27$, $\Omega_{\Lambda}$=0.73, $\textit{w}=-1$). Residuals can also be calculated removing the correction for stretch,
\begin{equation}
m_{B}^{resid, no~s}=m_{B}^{max}-M_B - 5~log_{10}~D_L - 25  - \beta C
\end{equation}
These residuals can be used to look for correlations with the wavelength/velocities of NUV features. We apply a redshift cut of $z\geq0.02$ (to exclude SNe that are affected by host galaxy peculiar velocity components), which excludes seven SNe, leaving 11 SNe for which the Hubble residuals can be calculated (after the phase and $B-V$ colour cuts detailed in Section \ref{selection_mean}).

\section{Results I - Mean comparisons}
\label{results1}

This section details comparisons performed between the mean spectra described in Section \ref{selection_mean}. 
We seek to test
the claim that the NUV spectra evolve with redshift as suggested
by previous works, \citep[e.g.][]{sul09, coo11, fol12}, and to 
understand whether the UV dispersion originally discussed by
E08 is an intrinsic property of SNe\,Ia or connected to this claimed evolution. The impact of these trends
on the cosmological utility of SNe\,Ia can be gauged by examining
possible correlations between the NUV spectra and the photometric
residuals on the Hubble diagram. We compare low and intermediate-$z$ samples, along with comparisons of the \textit{HST} SNe split based on stretch and Hubble residuals. The spectra split by $B-V$ colour have been described in Section \ref{colour_cor}, when discussing the colour corrections, and so are not described again here.

\subsection{Constructing mean spectra}
\label{selection_mean}

To look for any potential evolution with redshift in the SNe Ia sample, the new low-$z$ sample must be compared to a sample of SNe Ia NUV spectra at higher $z$ for which we use the intermediate-$z$ sample of E08. The spectra from this intermediate-$z$ sample have been analysed in the same manner as our low-$z$ sample, and we have colour-corrected them using the CCM colour law, as described in Section \ref{colour_cor}. For consistency, we have also refitted their light curves using the same version of SiFTO that is used for our low-$z$ sample.  \cite{gon10} investigated the $B-V$ colour-stretch relation of the SNLS SN Ia sample and found no correlation between light curve stretch and $B-V$ colour for `normal' SNe Ia. We also find no correlation between stretch and $B-V$ colour for our low-$z$ sample.

The mean spectra are created by rebinning the input spectra to 3 \AA\ per pixel, and then normalising the spectra through a box filter over a wavelength range 4000--4500 \AA. A three-sigma clipped mean is used and the error on the mean is determined by bootstrap resampling. In the mean plots shown here, the dashed lines refer to lower and upper 90 per cent confidence limits. The results are not changed by varying the normalisation region in the optical wavelength region.

SN Ia spectra in a suitable phase range must be selected when constructing the NUV mean spectra so that phase-related effects do not unnecessarily influence the mean spectrum comparisons. We have chosen an effective phase range of -1.0 to +4.5 d for both the low-$z$ sample and the intermediate-$z$ sample from E08. \cite{coo11} chose an effective phase range of -0.32 to +4 d, similar to our chosen values. We have also placed cuts on the stretch ($0.7<s<1.3$), and $B-V$ colour ($-0.25<c<0.25$) of the SNe in both samples. These are the same cuts as applied in the cosmological analysis of \cite{con11}. No SNe Ia in our low-$z$ sample are removed by either of the stretch cuts or the lower $B-V$ colour or phase cut. We lose nine SNe from the low-$z$ sample due to the upper phase cut and one SN (PTF10ygu) because of the upper $B-V$ colour cut. SN 2011ek is also excluded, since it displays an unusual NUV-$B$ colour of 1.0 mag, which will be discussed in Section \ref{nuvopty}.  This leaves a total of 17 SNe to be used in the mean low-$z$ spectrum. The preliminary low-$z$ mean spectrum of \cite{coo11} contained 10 \textit{HST} SNe Ia and one historical SN Ia from the literature. From the initial intermediate-$z$ sample of 33, 17 SNe are removed due to the phase cuts and one SN due to the $B-V$ colour cuts, leaving 15 SNe for the intermediate-$z$ mean. 

 Choosing SN Ia spectra in a suitably small effective phase range, yet that still contains a statistically significant number of objects, will limit the effect of phase variations when making comparisons between mean spectra. Despite this, phase variations within this range will still occur; for example, SNe cool with time, and their line velocities and NUV flux decrease. However, the effect of phase variations can be minimised by ensuring that when comparing mean spectra, the phase distributions (along with the stretch, $B-V$ colour and host galaxy distributions) of the samples are drawn from the same parent populations. As described in Section \ref{sec:sample-selection}, K-S tests were performed and for all these properties, there is no indication they are drawn from different parent populations.

\subsection{Evolution in the mean SN Ia spectrum with redshift}
\label{evo_red}

The mean spectrum at low-$z$ is compared to that of the intermediate-$z$ Keck SN Ia sample from E08 in Fig.~\ref{mean_highlow}. This is the first time that a large sample of low-$z$ SN Ia NUV spectra with colour corrections applied has been available for comparison with higher-$z$ samples. As described in Section \ref{selection_mean}, the effective phase, stretch, $B-V$ colour and host M$_{\rm stellar}$ distributions of the two samples are found to be well-matched. The mean spectra of the two samples are seen mainly to agree with the uncertainties (90 per cent confidence limits). However, at shorter wavelengths ($<3300$ \AA), the low-$z$ mean spectrum has less flux than the intermediate-$z$ sample.

To investigate this further, a random number of spectra (8-15) were used to make the low- and intermediate-$z$ mean spectra and this was repeated 20000 times. We compute the flux through a `UV' box filter, in the region 2900--3300 \AA, for each spectrum, and find that 99.8 per cent of the time (3.1-$\sigma$), the intermediate-$z$ mean spectrum has a greater flux than the low-$z$ mean spectrum.   \cite{coo11} also found a lower flux at shorter NUV wavelengths when compared to the sample of E08. The choice of this wavelength range of 2900-3300 \AA\ was motivated by the results of \cite{wal12}, which showed using radiative transfer modelling that this is the region where the most variation due to metallicity effects is expected. The importance of this flux evolution with redshift will be discussed in Section \ref{nuvopty}. 

We also compare in Fig.~\ref{mean_highlow} the dispersion in the mean low- and intermediate-$z$ SN samples. The dispersion is calculated as the flux error as a percentage of flux. There is an increased dispersion at wavelengths $<3700$ \AA, as found in previous studies. A peaked feature is visible at $\sim$3500--3800 \AA\ in the dispersion of both samples. This is found to be caused by an increase in the bootstrap resampling error in the lower part of the blue wing and the minimum of the \Caii\ H\&K feature and the relatively small flux in this region. 

We investigated the effect of host galaxy contamination on the results obtained for the mean spectrum of both the low- and intermediate-$z$ sample. As described in Section \ref{uv_spec}, the \textit{HST} sample is free from host galaxy contamination due to the narrow slits used to obtain the spectra. The host galaxy contamination of the intermediate-$z$ sample was estimated in E08 and the continuum flux levels in the NUV were found to be well constrained using the multi-band $u^*g'r'i'z'$ SNLS photometry. The removal of host galaxy emission lines from the spectra of this sample was more complicated and is the cause of the discrepancy in the trough of the \Caii\ H\&K feature, but this does not affect the overall flux levels. The effect of the colour correction used on mean spectra is also investigated, and the flux offset between the low- and intermediate-$z$ samples remains when different colour laws are used. 

\begin{figure*}
\includegraphics[width=18cm]{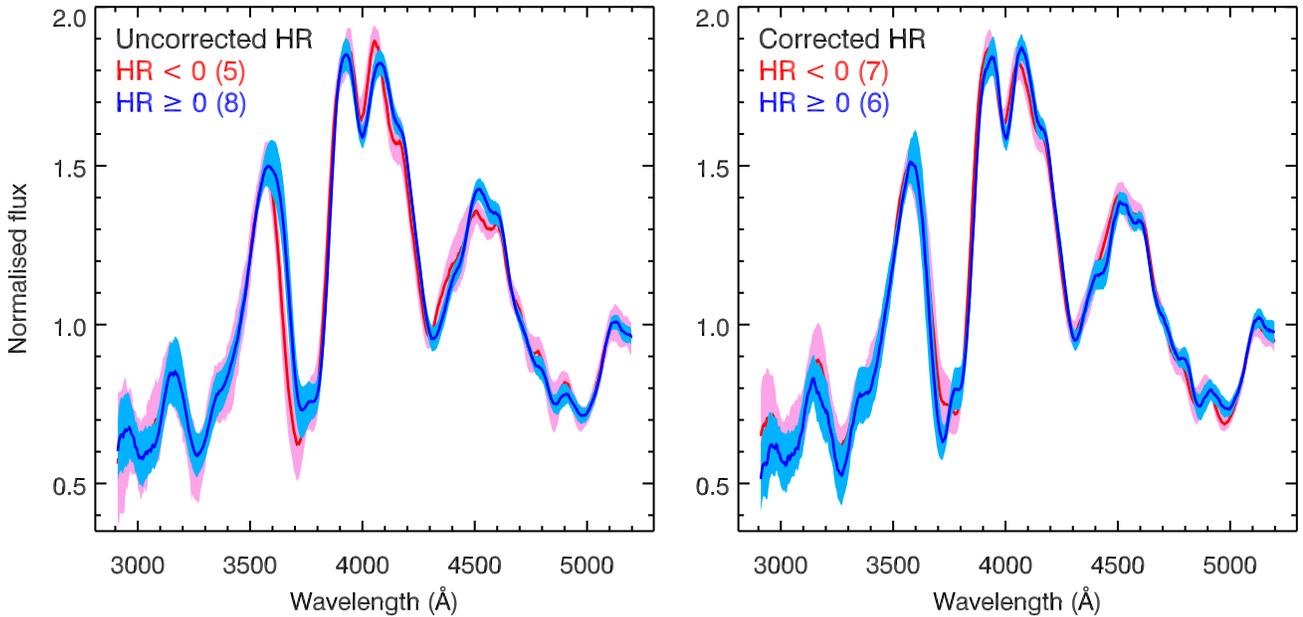}
\caption{The mean spectra of \textit{HST} SN Ia subsamples split by uncorrected Hubble residual value (left) and by corrected Hubble residual value (right). The division value is a Hubble residual of zero. The numbers in parentheses are the number of SN Ia spectra that went into each mean spectrum.}
\label{hub_mean}
\end{figure*}

\begin{figure*}
\includegraphics[width=18cm]{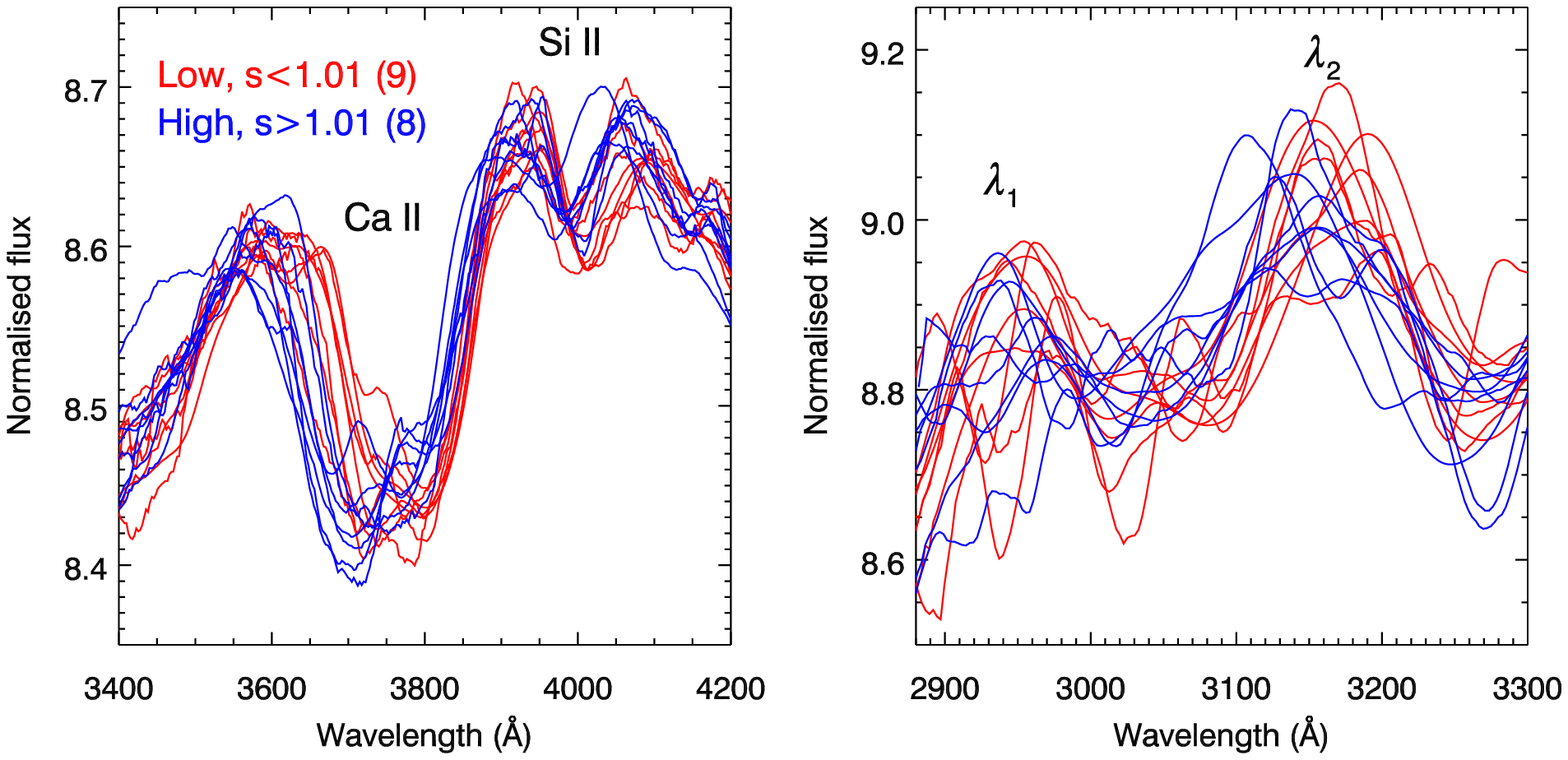}
\caption{The individual SN Ia spectra that are used in producing the mean spectra presented in Section \ref{results1}, are shown for the \Caii\ H\&K and \SiII\ 4130 \AA\ features on the left and $\lambda_1$, $\lambda_2$ features on the right. The SNe spectra are also split into high-stretch (blue) and low-stretch (red) bins. The numbers in the legends represent the number of SN spectra plotted for each stretch bin. The spectra have been smoothed with a boxcar average of 11 pixels. The spectra have been normalised in the region 3400-4300 \AA\ for the left panel and 3000-3400 \AA\ for the right panel.}
\label{cahk_str_over}
\end{figure*}

\subsection{Mean spectra split by light curve shape}
\label{stretch_mean}

Our sample can be split in various different ways to investigate possible astrophysical effects in SNe Ia. One of the most obvious is to split the sample according to the SN light curve stretch. We have split the sample into bins at the position of the mean stretch ($s=1.01$) of the sample. This results in 9 and 8 SNe in the low- and high-stretch bins respectively. Fig.~\ref{mean_stretch2} shows a comparison of the low- and high-stretch mean spectra for the low-$z$ sample. A blue-shift in the wavelength positions of the NUV features is clearly seen for the higher stretch SNe. In particular, the blue wings of \Caii\ H\&K feature at $\sim$3700 \AA\ and the NUV features, $\lambda_1$ and $\lambda_2$ at  2920 and 3180 \AA\ respectively, are blue-shifted in high-stretch compared to low-stretch SNe Ia. We have shown that the phase distributions of the low- and high- stretch samples are well-matched (97 per cent probability they are from the same parent populations). Therefore, the effect of different phase distributions can be ruled out as the cause of the velocity shift seen in Fig.~\ref{mean_stretch2}. 

Fig.~\ref{mean_stretch2} shows that the low-stretch SNe in our sample have slightly more NUV flux at the positions of the  $\lambda_1$ and $\lambda_2$ features. As will be discussed in Section \ref{discussion}, this is most likely linked to the velocity differences seen in the NUV, with higher velocities causing more line blanketing and hence, a greater suppression of the NUV flux for high, compared to low velocity events.  

The dispersion of the two stretch samples is also shown in Figure \ref{mean_stretch2}, where an increased dispersion towards shorter wavelengths is observed. This is similar to the increased dispersion found in both the low and intermediate-$z$ mean spectra. Similarly to the spectra split by redshift, a peaked feature is present at a slightly shorter wavelength than $\sim3700$ \AA, when the sample is split by stretch. The peak of the high-stretch dispersion is shifted by 21 \AA\ with respect to the low-stretch dispersion. This feature is caused by a combination of an increased bootstrap resampling error at the minima and lower half of the blue-wings of the \Caii\ H\&K features in both spectra, and the smaller flux in this region than at the blue-edge continuum of the feature.

\subsection{Mean spectra split by Hubble residuals}
\label{sec_mean_hub}

One interesting parameter upon which our sample can be split is Hubble residual values. These values were calculated as described in Section \ref{make_HR}. In Fig.~\ref{hub_mean} the mean spectra split by uncorrected Hubble residual and by corrected Hubble residual are shown. The uncorrected Hubble residuals are the values before corrections for either light curve stretch or colour are applied (see Section \ref{make_HR}). The corrected Hubble residuals are those after these corrections have been applied. The velocity shift in the blue wing of the \Caii\ H\&K feature for SNe Ia with negative Hubble residuals seen in the left panel of Fig.~\ref{hub_mean} is in agreement with the results of the mean spectra split by stretch. There is good agreement between the two spectral comparisons shown in Fig.~\ref{hub_mean}, with no obvious differences between the positive and negative Hubble residual samples.

\section{Results II - Analysis of individual spectra}
\label{results2}
Although the comparison of mean spectra is a useful tool for assessing evolution in SN properties as a whole, the analysis of individual spectra can be used to investigate colour variations, along with correlations between spectral properties and other SN Ia properties. Individual line measurements are also advantageous over mean spectrum comparisons because it is possible to correct for phase variations, which can occur even within the narrow phase window used for the mean comparisons (-1.0--4.5 d). The same cuts on the sample as in Section \ref{results1} are used.

 The NUV colours can be calculated from the spectra and compared with the $B-V$ colour obtained from the light curve fits to look for increased dispersion at NUV wavelengths. An analysis of the features in individual spectra is also performed, including the wavelength shifts of the \SiII\ 4130 \AA, \Caii\ H\&K, $\lambda_2$ and $\lambda_1$ features. The $\lambda_1$ and $\lambda_2$ features are defined as the NUV peaks at $\sim$2920 and $\sim$3180 \AA\ respectively as shown in Figure \ref{mean_highlow}. These features are suggested from modelling to be dependent on the metallicity of the progenitor and its composition \cite[e.g.][]{len00, sau08, wal12}.  Therefore, a shift in these features may be driven by the properties of the progenitor system.

In Fig.~\ref{cahk_str_over}, the individual spectra that are used in the measurements presented in this section, are shown for the \Caii\ H\&K and \SiII\ 4130 \AA\ features on the left, and $\lambda_1$, $\lambda_2$ features on the right. The SNe spectra are further split based on their stretch, into low- and high-stretch bins at a position of $s=1.01$.  It can be seen clearly that there is a stretch-dependent shift in the wavelength position of the \Caii\ H\&K, $\lambda_2$ and \SiII\ 4130 \AA\ features, with higher stretch SNe having greater velocities, similar to the results of Section \ref{stretch_mean}. These trends between the wavelength shift of the features and stretch will be investigated further in the rest of this section. The increased dispersion at shorter wavelengths is also clearly visible in Fig.~\ref{cahk_str_over}, where there is a much larger variation in the positions of the $\lambda_2$, and particularly $\lambda_1$ features compared to the relative homogeneity (apart from velocity differences) of the \Caii\ H\&K feature. This increased dispersion will investigated further using synthetic filters in Section \ref{nuvopty}.

\begin{figure*}
\includegraphics[width=18cm]{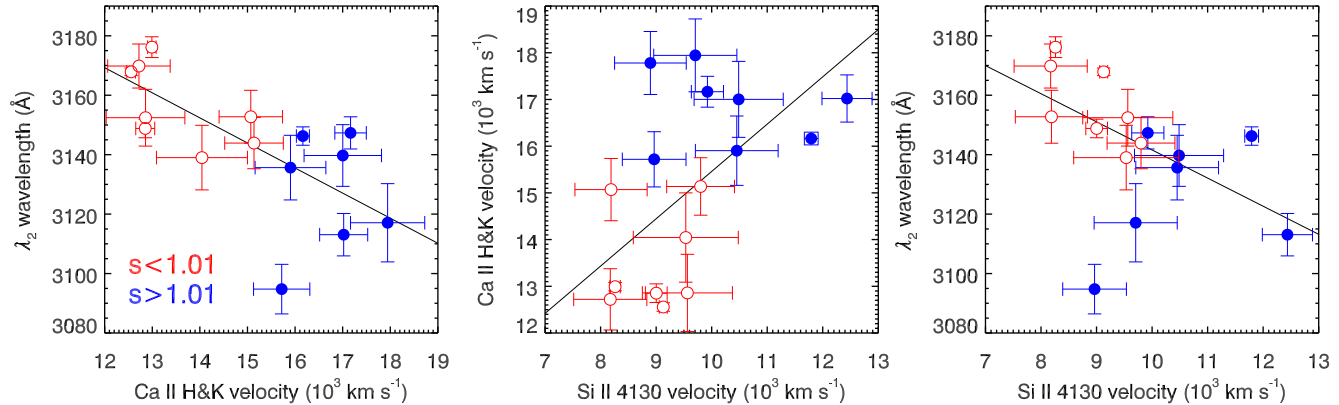}
\caption{The measured velocities (corrected for phase variations) of \Caii\ H\&K and  \SiII\ 4130, along with the wavelength of the $\lambda_2$ feature are plotted as functions of each other. The left panel is $\lambda_2$ against \Caii\ H\&K, the middle panel is
 \Caii\ H\&K plotted against \SiII\ 4130 \AA\ and the right panel is $\lambda_2$ against \SiII\ 4130 \AA. The high-stretch ($s>1.01$) and low-stretch ($s<1.01$) samples have been displayed in blue filled circles and red open circles respectively. There is one less SN in the high-stretch bin for the plots containing $\lambda_2$, because the $\lambda_2$ feature could not be identified for one SN (PTF09dlc). The best fit lines to the data in each panel are plotted as solid, black lines. The significances of a non-zero gradient are 3.4, 2.8 and 2.2-$\sigma$ from left to right.}
\label{si_ca_lam}
\end{figure*}

\subsection{Line measurements}
\label{line_measure}
To measure the position of the minima and maxima of the various spectral features, we fit a Gaussian to the line profile. This is done by subtracting off the pseudo-continuum, then choosing maximum values on each side of the feature and performing a series of IDL's MPFIT \citep{mar08} fits around this initial value. The positions of the pseudo-continuum points are then varied within a range of $\pm$30 \AA\ and the minimum wavelength is chosen as the minimum of the fit with the lowest $\chi^2$ value. The wavelength position errors are estimated from the range in these fit values. The total uncertainties of the wavelength measurements include the uncertainty in measuring the position, selecting the pseudo-continuum and the redshift uncertainties. All the spectra were visually inspected to ensure the fit and the chosen pseudo-continuum values are sensible.

\cite{fol12} (whose data will be compared to ours in Section \ref{cahk_vel}) measure the line position of the \Caii\ H\&K feature following the method of \cite{blo06} and the culling criteria of \cite{fol11}. This method involves smoothing the spectra using an inverse-variance-weighted Gaussian filter. The smoothed spectra are resampled onto a fine wavelength scale of 0.1 \AA\ and the minima of the features are measured. If two minima are identified in the \Caii\ H\&K feature, they are classified separately as ``blue'' or ``red'' components. Five SN spectra present in the Keck sample are also part of the \cite{fol12} sample. We compare our measurements of the \Caii\ H\&K velocities to those presented in \cite{fol12} and find our measured values are lower than the values of \cite{fol12} by values ranging from $\sim$650--3800 \kms. \cite{bro08} also used a Gaussian fit to the entire feature, while \cite{sil12b} fitted the entire \Caii\ H\&K line profile using a cubic spline to identify the position of the minimum.

We have chosen not to follow the method of \cite{fol12} since the presence of additional features present in the \Caii\ H\&K features could bias the measurements of these parameters. Therefore, we fit a Gaussian to the overall \Caii\ H\&K feature. Although this will include a contribution from the \SiII\ feature, which has been identified in the \Caii\ H\&K feature using the spectral fitting code, \textsc{synapps}, the contribution from \SiII\ can be estimated using the velocity outputs from \textsc{synapps} (see Section \ref{cahk_vel}).

\subsection{Correction for phase variation}
\label{phase}

As for our mean spectrum comparisons in Section \ref{results1}, we measure wavelength positions/velocities for the studied features (\Caii\ H\&K, $\lambda_1$, $\lambda_2$ and \SiII\ 4130 \AA) in the SN spectra in the effective phase range of -1.0 to +4.5 d. However, variations as a function of time within this chosen phase range may still be present and must be corrected for. We correct these variations by fitting the wavelength positions/velocities as a function of phase for our SN samples and adjusting to a phase of 0 d. We simultaneously fit the values of both the \textit{HST} and Keck spectra for the \Caii\ H\&K velocity, $\lambda_2$ wavelength and \SiII\ 4130 \AA\ velocity, while for the $\lambda_1$ feature, we fit just the \textit{HST} spectra due to their higher signal-to-noise (S/N). The fits are performed using equations (depending on whether wavelength or velocity is being corrected) of the form (here for velocity),

\begin{equation}
v_0=v_t - \dot{v}~t
\end{equation}
where $v_t$ is the velocity (\kms) measured from the spectra, $v_0$ is the velocity (\kms) at maximum, $t$ is the effective phase (d) of the spectrum and $\dot{v}$ is gradient of the fit ( \kms d$^{-1}$). The values of the gradient of the fits are shown in Table \ref{phase_corr}, along with the maximum correction to 0 d that was applied. The NUV$-$optical colours measured from the NUV spectra are also corrected for phase variations following the same method.  For the NUV$-$optical colour measurements, we fit the values of only the \textit{HST} sample. The measured (and phase-corrected) wavelengths and velocities are given in Table \ref{vel_spec}.

Ideally the velocity evolution of the individual SNe Ia should be studied but given the single epoch spectra available for our sample, this was not attempted.  However, a decrease in velocity/increase in wavelength is seen as a function of time for our sample, as has been found using measurements of individual SN spectra as a function of time \cite[e.g.][]{hac06, fol11, sil12b}. Therefore, on average, our corrections should remove this phase dependence in the range of -1.0 to +4.5 d. Our NUV \textit{HST} Cycle 18 data, with multi-epoch spectra of 4 `normal' SNe Ia will provide improved measurements of the diversity of the phase evolution of NUV features for future studies.

\begin{table}
\caption{The gradient of the phase corrections that were used to correct the wavelengths of the studied feature and colour to a phase of $B$-band maximum, and size of the correction applied.}
\begin{center}
\begin{tabular}{|lll|llll}
\hline
\hline
Parameter & Gradient & Size of \\
&&correction \\
\hline
\Caii\ H\&K &$-$280$\pm$230 \kms d$^{-1}$&$<1280$   \kms  \\
$\lambda_2$ &6.12$\pm$2.61 \AA\ d$^{-1}$&$<25$  \AA\\
$\lambda_1$ &$-$1.64$\pm$2.75 \AA\ d$^{-1}$&$<7$  \AA \\
\SiII\ 4130 & $-$91$\pm$60 \kms d$^{-1}$& $<372$   \kms\\
$UV-b$&0.06$\pm$0.02 mag d$^{-1}$&$<0.3$ mag\\
\end{tabular}
\end{center}
\label{phase_corr}
\end{table}%

\subsection{\Caii\ H\&K, \SiII\ 4130 \AA\ and $\lambda_2$ positions}
\label{all_plot}
Fig.~\ref{si_ca_lam} shows the wavelength positions of the \Caii\ H\&K, \SiII\ 4130 and $\lambda_2$ features plotted as functions of each other. The left panel is $\lambda_2$ against \Caii\ H\&K, the middle plot is \Caii\ H\&K plotted against \SiII\ 4130 \AA, and the left plot panel is $\lambda_2$ against \SiII\ 4130 \AA. The high- and low-stretch SN samples are also shown, split at a value of $s = 1.01$. The best fit lines to the data for each comparison are also measured and have significances from zero of 3.4, 2.8 and 2.2-$\sigma$ for  $\lambda_2$ against \Caii\ H\&K, \Caii\ H\&K plotted against \SiII\ 4130 \AA\ and $\lambda_2$ against \SiII\ 4130 \AA, respectively.

The \SiII\ 4130 \AA\ absorption feature has been previously considered a good measure of the photospheric velocity and there is a correlation between this feature and the \Caii\ H\&K and $\lambda_2$ features. However, the \SiII\ 4130 \AA\ feature is investigated using \textsc{synapps} and it is found that \SiiNF\ and \Coii\ are found to contribute in this region. This could explain the lack of a stronger correlation between the measurement of the \SiII\ 4130 \AA\ `notch'  and the \Caii\ H\&K velocity seen in Fig.~\ref{si_ca_lam}, since this is not measuring \SiII\ alone. Further investigation of the NUV features using both correlations with photometric properties, and using \textsc{synapps} spectral fitting will be presented in the following sections.

\begin{figure}
\includegraphics[width=9cm]{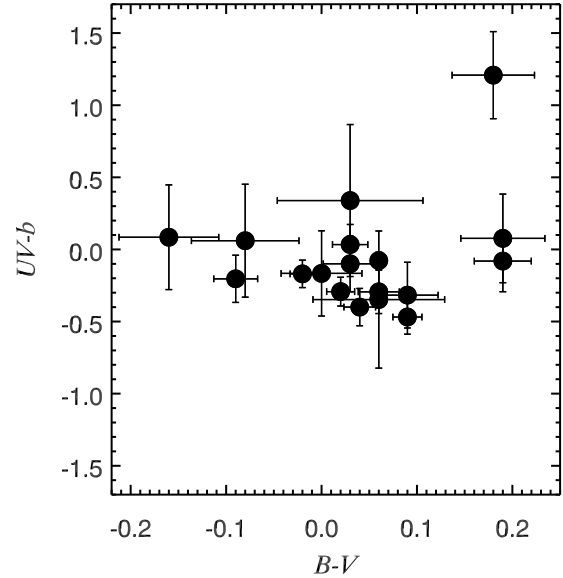}
\caption{The $UV-b$ colour (after phase and SiFTO $B-V$ colour correction)  is shown as a function of $B-V$ colour. The outlier in the plot is SN 2011ek, which has a very red $UV-b$, along with one of the reddest $B-V$ colour values. }
\label{ratio_uv1b_str_col}
\end{figure}

\subsection{NUV-optical flux comparison}
\label{nuvopty}
The evolution with redshift in the NUV spectral features, as well as the increased dispersion at NUV wavelengths shown in Section \ref{evo_red} can also be investigated using synthetic filters. As discussed in Section \ref{evo_red}, to investigate the significance of the observed evolution with redshift, we define a `UV' box filter between 2900--3300 \AA, since this is the spectral region which is shown by \cite{wal12} to have the largest metallicity variations. This `UV' filter is 100 \AA\ redward of the `UV1' filter used in E08. We compare the flux through this filter to that in a `b'-band filter with a wavelength range of 4000-4800 \AA. 
 
We investigated the size of the evolution with redshift in the mean spectra by comparing the mean weighted $UV-b$ colours of the low- and intermediate-$z$ samples, and find that they have values of -0.18$\pm$0.03 and -0.36$\pm$0.04 mag, respectively. This difference of 0.18 mag (3.6-$\sigma$ effect) is in agreement with the results of the mean spectrum comparison at different redshifts, shown in Section \ref{evo_red}, where the low-$z$ is redder at NUV wavelengths compared to the intermediate-$z$ sample. We discuss the significance of this in Section \ref{discussion}. 

Fig.~\ref{ratio_uv1b_str_col} shows the measured $UV-b$ colours as a function of $B-V$ colour (after applying the CCM-law colour correction and the phase correction given in Table \ref{phase_corr}). The same phase and $B-V$ colour cuts as used for creating the mean spectra are used for this sample. SN 2011ek was excluded from the mean spectrum analysis because of its unusually red $UV-b$ colour that is a clear outlier ($UV-b$ of 1.0 mag) when compared to the SNe shown in Fig.~\ref{ratio_uv1b_str_col}.  We find no correlation between the $UV-b$ colour and stretch. However, the scatter in $UV-b$ is significantly larger than in $B-V$, with an r.m.s. of 0.21 versus 0.08 mag (excluding SN 2011ek). This is consistent with previous studies that noted a larger dispersion in NUV spectra compared to the optical \cite[e.g. E08;][]{fol08a}. Again, we discuss the possible origin of these effects in Section \ref{discussion}.

\begin{figure*}
\includegraphics[width=18cm]{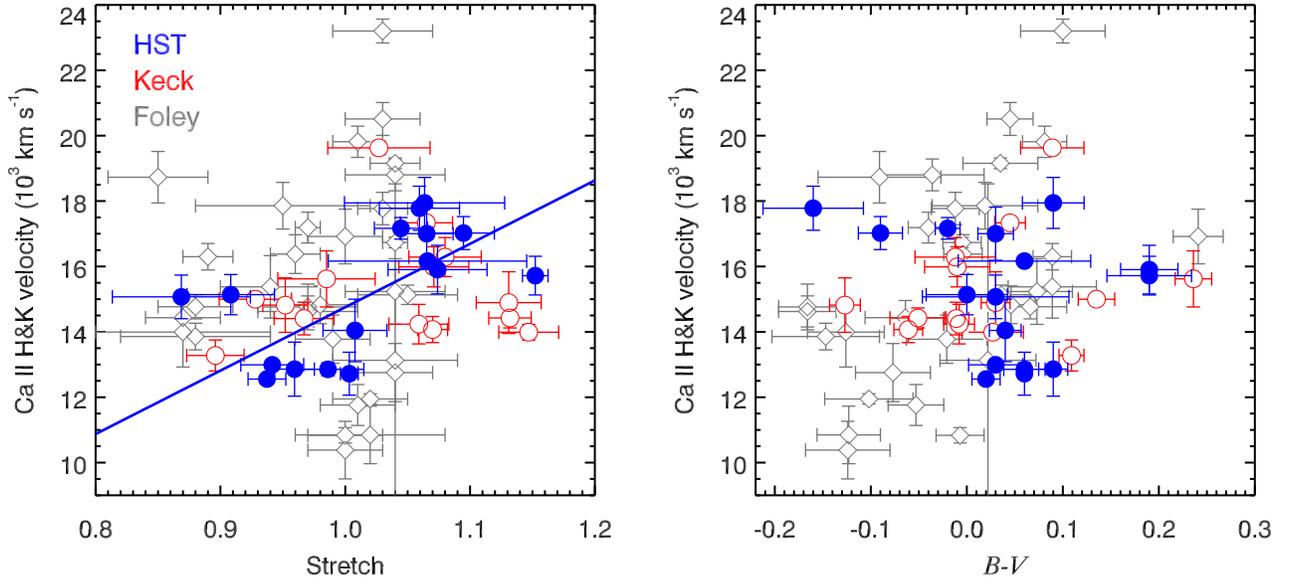}
\caption{The wavelength position of the \Caii\ H\&K feature for the \textit{HST} (blue, solid circles) and Keck (red open circles) samples against stretch is shown in the left panel, while the right panel shows the velocity against $B-V$ colour.  The \Caii\ H\&K velocities have been phase-corrected using the relation in Section \ref{phase}. The grey, open diamonds are the measured  \Caii\ H\&K values from  \protect \cite{fol12}, which we have corrected using the same relationship as for the \textit{HST} and Keck data. The solid blue line is the best fit to the \textit{HST} data with a significance of 3.4-$\sigma$ from zero. The same sample cuts as for the \textit{HST} and Keck samples have been  also applied to the \protect \cite{fol12} data.  }
\label{cahk_str_col}
\end{figure*}

\begin{figure*}
\includegraphics[width=18cm]{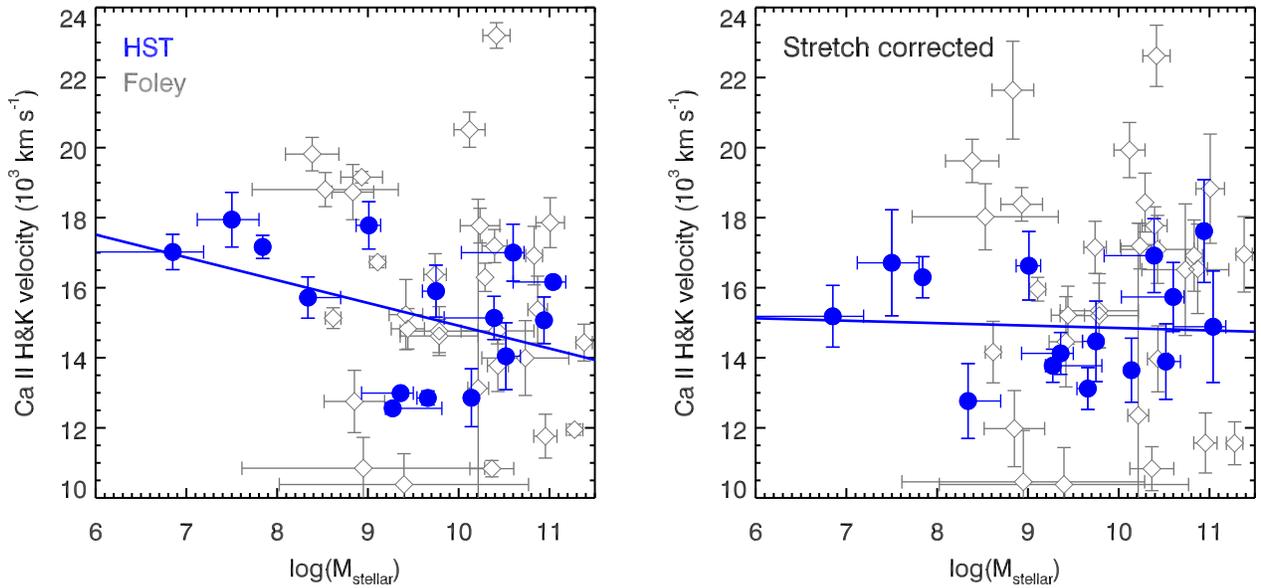}
\caption{The measured velocities (corrected for phase only) of the \Caii\ H\&K feature for the \textit{HST} sample and the sample from \protect \cite{fol12} are shown as a function of log(M$_{\rm{stellar}}$) in the left panel. The solid blue line is a linear fit to the \textit{HST} data only, showing a correlation between \Caii\ H\&K velocity and host galaxy M$_{\rm{stellar}}$ (1.7-$\sigma$ from zero). \protect \cite{fol12} also identified a similar trend in their data. The right panel shows the \Caii\ H\&K velocities for the samples after the correlation between \Caii\ H\&K velocity and light curve stretch has been removed. A linear fit to the \textit{HST} data is plotted as a solid blue line and the correlation between \Caii\ H\&K velocity and host galaxy M$_{\rm{stellar}}$ is seen to be removed.}
\label{cahk_mass_vel}
\end{figure*}

\subsection{\Caii\ H\&K velocity}
\label{cahk_vel}

The \Caii\ H\&K wavelength positions for both the \textit{HST} and Keck sample were measured and corrected for phase variations as detailed in Sections \ref{line_measure} and \ref{phase}. These wavelengths were then converted to velocities using the relativistic Doppler formula and a rest wavelength taken as 3945 \AA\ (the gf-weighted average of the 3933, 3969 \AA\ features). The velocities of the \Caii\ H\&K feature for the \textit{HST} and Keck samples are shown in Fig.~\ref{cahk_str_col} as a function of stretch in the left panel and $B-V$ colour in the right panel. \cite{fol12} have also measured the \Caii\ H\&K velocity of a sample of SDSS and SNLS SNe Ia. Their measured values which are within our chosen phase range are corrected to maximum light using our phase relationship and are also plotted in Fig.~\ref{cahk_str_col}. No correlation between stretch and $B-V$ colour is apparent in our sample, in agreement with previous larger studies \cite[e.g.][]{bra10, gon10}.

A trend of increasing velocity with increasing stretch is seen in the \textit{HST} and Keck samples. A linear fit was performed to the \textit{HST} sample and is shown in Fig.~\ref{cahk_str_col}. The fit has a significance from zero of 3.4-$\sigma$.  This higher velocity with stretch trend seen for the \textit{HST} sample is consistent with Fig.~\ref{mean_stretch2}, where the \Caii\ H\&K feature is shifted to the blue for the higher stretch SNe.
 A contribution to the \Caii\ H\&K region from \SiII\ is identified using the spectral fitting code, \textsc{synapps}. The \SiII\ component is weaker in all cases than the contribution from \Caii. The identified trend in the observed data between \Caii\ H\&K velocity and stretch can be investigated using the velocity outputs from \textsc{synapps}. A correlation between the fitted \Caii\ velocity and stretch is found at a significance of 2.4-$\sigma$ from zero. The velocity of the \SiII\ component is also found to correlate with stretch (2.3-$\sigma$). This suggests that the identified trend of increasing velocity of the overall \Caii\ H\&K features with increasing stretch (luminosity), is a result of a trend in both in the \Caii\ and \SiII\ components.

 In the right panel of Fig.~\ref{cahk_str_col}, the \Caii\ H\&K velocities as a function of $B-V$ colour are shown, for the \textit{HST}, Keck and Foley samples. No trend with $B-V$ colour is identified in either the \textit{HST} or Keck samples, while for the sample of \cite{fol12}, their identified trend of increasing colour with increasing \Caii\ H\&K velocity is clearly seen in their data. \cite{sil12} also found no trend between \Caii\ H\&K velocity and $B-V$ colour for their low-$z$ sample.

The sample of \cite{fol12} includes only SNe Ia with s$<1.05$, while the \textit{HST} data contains 7 SNe with stretch values greater than this. When these high-stretch SNe are excluded from our low-$z$ sample, the trend between \Caii\ H\&K and stretch drops to less than 1-$\sigma$ from zero. Analysis of the full range of stretch values used in current cosmological samples demonstrates the trend of increasing \Caii\ H\&K velocity with stretch. The reason for the observed discrepancies between \Caii\ H\&K velocity and $B-V$ colour, when comparing the \textit{HST} sample, to that of \cite{fol12} is unclear. A possible reason may be the differences in the velocity measurement techniques used, as discussed in Section \ref{line_measure} or the definition of $B-V$ colour. 

The host galaxy properties of SNe Ia have been demonstrated to have a significant impact on the properties of the SNe they host. We investigate here if there is a relation between host galaxy M$_{\rm{stellar}}$ (a good proxy for metallicity) and \Caii\ H\&K velocity. We find a correlation between these two parameters for the \textit{HST} sample at a significance of 1.7-$\sigma$ from zero. This is in agreement with the results of \cite{fol12}, who also find a similar correlation in their sample. The left panel of Fig.~\ref{cahk_mass_vel} demonstrates this relation, showing the velocity of the \Caii\ H\&K line of the \textit{HST} and the sample of \cite{fol12} as a function of the log(M$_{\rm{stellar}}$). This correlation also agrees with the relation between \Caii\ H\&K and stretch seen in Fig.~\ref{cahk_str_col}, since it has been shown that less massive, star-forming galaxies preferentially host SNe Ia with broader, brighter light curves \citep{ham96, ham00, sul06}. From Fig.~\ref{cahk_mass_vel}, it can also be seen that there is a lack of low velocity (low stretch) SNe in hosts with low M$_{\rm{stellar}}$.

To investigate if the \Caii\ H\&K velocity and M$_{\rm{stellar}}$ correlation is caused by the previously identified stretch and M$_{\rm{stellar}}$ correlation, we remove the observed \Caii\ H\&K velocity and stretch correlation shown. In the right panel of Fig.~\ref{cahk_mass_vel}, the \Caii\ H\&K velocity (after correlation for the \Caii\ H\&K velocity versus stretch has been removed) as a function of M$_{\rm{stellar}}$ is shown, and the relation between \Caii\ H\&K velocity and M$_{\rm{stellar}}$ is found to be removed. This shows that the observed \Caii\ H\&K velocity versus M$_{\rm{stellar}}$ correlation is driven by the well-known correlation between host galaxy M$_{\rm{stellar}}$ and light curve stretch.

\subsection{Wavelength of NUV $\lambda_2$ feature}
\label{l2}
\begin{figure}
\includegraphics[width=9cm]{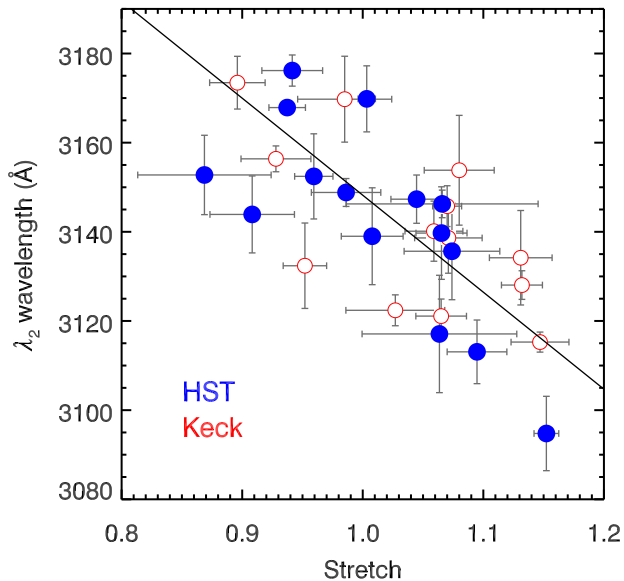}
\caption{The wavelength position of the $\lambda_2$ feature for the \textit{HST} and Keck samples, as a function of stretch is shown. A best linear fit to both data sets is plotted as a solid, black line in the left panel and has a significance of  6.2-$\sigma$. A linear fit to the \textit{HST} sample alone has a significance of 3.7-$\sigma$.}
\label{lam2_pos_colstr}
\end{figure}

The emission peak corresponding to the $\lambda_2$ was chosen by selecting the closest feature to the $\lambda_2$ of 3180 \AA\ as defined in E08 and is shown in Fig~\ref{mean_highlow}. Fig.~\ref{lam2_pos_colstr} shows the position of the $\lambda_2$ feature as a function of stretch.  A clear correlation between the position of the $\lambda_2$ feature and stretch is seen: lower stretch SNe Ia have a redder wavelength position of the $\lambda_2$ feature than higher stretch SNe. A best linear fit is determined with a significance of 6.2-$\sigma$ from zero for the combined \textit{HST} and Keck samples. Therefore, this trend is even stronger than that identified for the \Caii\ H\&K velocity. We find no trend between the $\lambda_2$ wavelength and $B-V$ colour. 

The position of the $\lambda_2$ feature as a function of M$_{\rm{stellar}}$ shows a trend of increasing blueshift with decreasing host $M_{\rm stellar}$, with a significance of 2.0-$\sigma$. More generally, there is a lack of SNe Ia with large $\lambda_2$ blue-shifts in high $M_{\rm stellar}$ galaxies, and a lack of events with small $\lambda_2$ blueshifts in low $M_{\rm stellar}$ galaxies. This is in agreement with the \Caii\ results. Similarly to the \Caii\ H\&K correlation, after correcting for the $\lambda_2$-stretch correction, the significance of the relationship between the wavelength of $\lambda_2$ and $M_{\rm stellar}$ drops to 0.5-$\sigma$.

 \subsection{Wavelengths of NUV $\lambda_1$ and \SiII\ 4130 \AA\ features}
\label{l1}

The position of the $\lambda_1$ feature (2920 \AA) was also measured. A weaker trend with stretch compared to $\lambda_2$ was found, but in the same sense. No trends with $B-V$ colour were found. The velocity of the \SiII\ 4130 \AA\ feature also correlates with stretch (higher stretch SNe have faster \SiII\ velocities) at 2.3-$\sigma$. As discussed in Section \ref{all_plot}, this feature is contaminated  by contributions from  \SiiNF\ and \Coii, and so may not be a clean tracer of the photospheric velocity. No trend between \SiII\ 4130 \AA\ velocity and $B-V$ colour is seen. We do not investigate the host galaxy nor Hubble residuals for these features since any trends are of lower significance, and instead focus our analysis on the \Caii\ H\&K and $\lambda_2$ features.

\subsection{Hubble residuals}
\label{HR_results}

\begin{figure*}
\includegraphics[width=18cm]{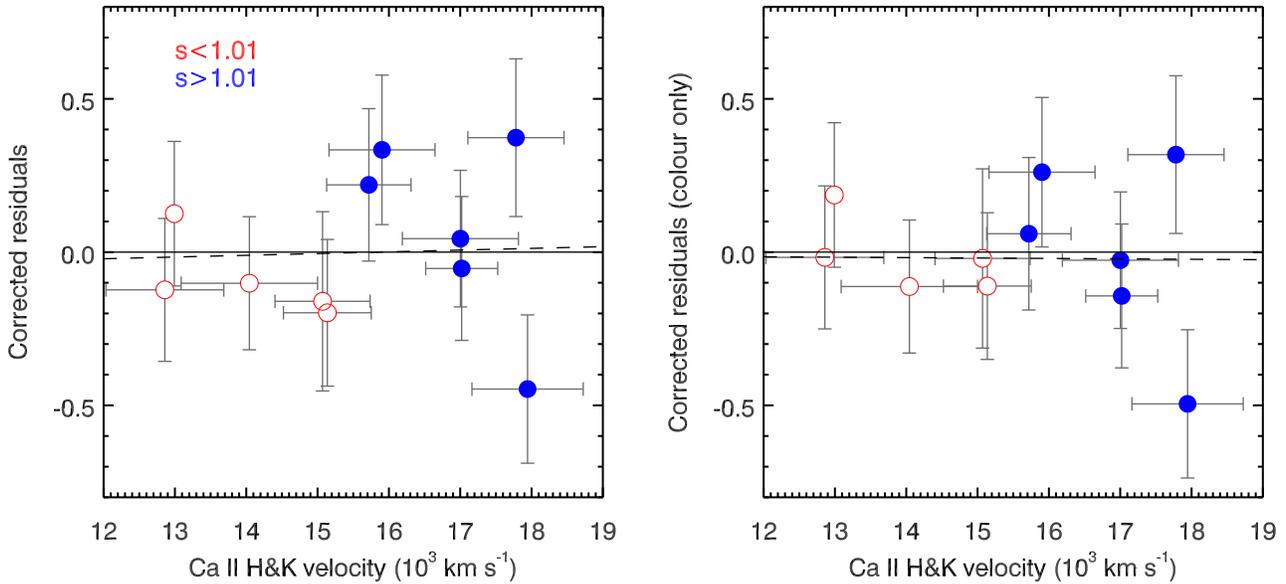}
\caption{The $B-V$ colour and stretch-corrected Hubble residuals and colour-only corrected Hubble residuals for the \textit{HST} sample as a function of \Caii\ H\&K velocity is shown in the left and right panels respectively. The SN measurements are also split based on their stretch into low-stretch (red, open circles) and high-stretch (blue, solid circles). The dashed lines are the best-fitting lines to the data.}
\label{cahk_hub_resid}
\end{figure*}

\begin{figure*}
\includegraphics[width=18cm]{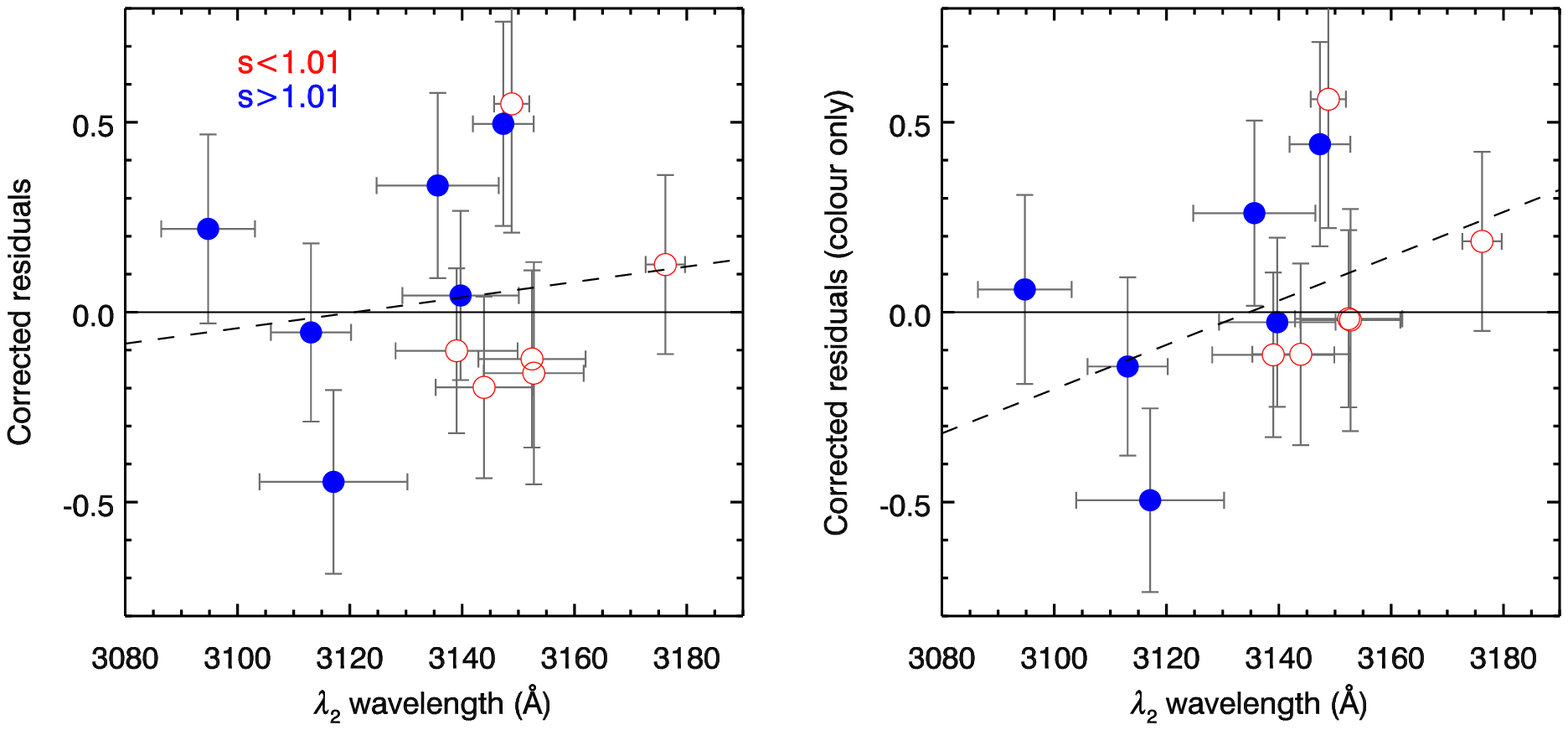}
\caption{The $B-V$ colour and stretch-corrected Hubble residuals and colour-only corrected Hubble residuals for the \textit{HST} sample are shown as a function of $\lambda_2$ in the left and right panels respectively. The SN measurements are also split based on their stretch into low-stretch (red, open circles) and high-stretch (blue, solid circles). The dashed line in the right panel is the best linear fit to the data with a significance of 1.7-$\sigma$ from zero.}
\label{lam2_pos_hub}
\end{figure*}

The Hubble residuals $m_{B}^{resid}$ (with $B-V$ colour and stretch corrections) and $m_{B}^{resid, no~s}$ (with $B-V$ colour, but no stretch correction) are shown in Fig.~\ref{cahk_hub_resid} as a function of the wavelength of the \Caii\ H\&K feature for the \textit{HST} sample. No trends between $m_{B}^{resid}$ or $m_{B}^{resid, no~s}$ and \Caii\ H\&K velocity are identified in the data. This suggests that, at least for this small sample of 11 SNe, there is no additional correlation between \Caii\ H\&K velocity and Hubble residual. Fig.~\ref{lam2_pos_hub} shows the Hubble residuals as a function of the $\lambda_2$ position. Again, for $m_{B}^{resid}$, no correlations are seen. However, a correlation with $m_{B}^{resid, no~s}$ as a function of $\lambda_2$ is apparent in the right panel of Fig.~\ref{lam2_pos_hub} but with low significance (1.7-$\sigma$). 
 
\section{Discussion}
\label{discussion}

In this paper, we have studied a large, low-$z$ sample of SN Ia NUV spectra obtained with \textit{HST}+STIS.  This sample contains 32 high S/N NUV spectra around maximum light, greatly improving upon the size of previous samples.
Photometric data were obtained for our sample using the LT and LCOGT robotic telescopes, so that light curve parameters could be calculated using the light-curve fitting routine, SiFTO. This allowed colour-correction of the \textit{HST} spectra to account for the effects of dust extinction and intrinsic colour variations, a correction that could not be applied in the earlier study of our dataset \citep{coo11}.

\subsection{Physical significance of an evolution with redshift and its dispersion}
In Section \ref{evo_red}, we compared our new low-$z$ mean spectrum to one constructed from the intermediate-$z$ sample of E08, and found a modest, but statistically significant (3.1-$\sigma$ effect), evolution with redshift at NUV wavelengths ($<$3300 \AA), with the low-$z$ mean spectrum having a depressed NUV flux compared with the intermediate-$z$ mean. This was confirmed using measurements of the $UV-b$ colour of the spectra in Section \ref{nuvopty}, where we showed that the mean $UV-b$ colour of the low-$z$ sample is 0.18 mag redder than that of the intermediate-$z$ sample. Previous studies including E08, \cite{coo11} and \cite{fol12b} also found a depressed NUV flux at low-$z$ compared to intermediate-$z$. However, this is the first time that an unbiased, large low-$z$ SN Ia sample has been used. We also find an increased spectral dispersion at NUV compared to optical wavelengths, confirming the results of previous studies such as E08.

Previous SN Ia modelling efforts have studied the effect of metallicity and composition on SN Ia properties \cite[e.g.][]{hof98, len00, sau08}. However, these studies have been unable to explain the size of the observed increase in dispersion at NUV compared to optical wavelengths in SNe Ia; they predicted much smaller variations than those observed in our NUV spectra. Some modelling efforts have focussed on asymmetries in the SN ejecta \citep{kro09}, where viewing angle effects were investigated as a potential cause of this dispersion. However, these models predict an effect of 0.5 mag in the $V$ band, which is much larger than the scatter seen in optical studies of SNe Ia.

However, the recent spectral synthesis models of \cite{wal12}, which vary the metallicity of the SNe, shed light on this observed evolution in NUV properties with redshift. Using a sequence of models at a fixed luminosity (log(L$_{\rm bol}$/L$_{\sun}$) = 9.6), where the metal content is varied by a factor of 0.05--5 times relative to that of SN 2005cf, we integrate their model spectra through our UV and $b$ filters, and find that the observed change with redshift in the $UV-b$ colour of 0.18 mag corresponds to a decrease in metallicity of 0.4 dex with increasing redshift.

Galaxy M$_{\rm stellar}$ is a crude proxy for both gas-phase and stellar metallicity \citep{tre04, gal05}. Our samples are matched in M$_{\rm stellar}$ (see Section \ref{sec:sample-selection}), however the relation between M$_{\rm stellar}$ and metallicity is expected and observed to evolve with redshift \cite[e.g.][]{sav05, lam09, cre12}. To determine the expected metallicity decrease with increasing redshift for our SN Ia samples, we firstly estimated the mean age of formation of the SNe at the different mean redshifts of the samples ($z=0$ and $0.6$) using the star formation history from \cite{li08} and the t$^{-1}$ delay-time distribution from \cite{mao11}. These redshifts correspond to mean formation redshifts of 0.3 and 0.8 for the low- and intermediate-$z$ samples, respectively. At these redshifts for fixed M$_{\rm stellar}$, we estimated a decrease in metallicity of $\sim$0.2 dex from \cite{lam09}. To first order, this is not too different to the metallicity evolution inferred based on the spectral modelling study of \cite{wal12}.  

These new SN Ia models also give insight into the origin of the $UV-b$ scatter. We again compare to the UV SN Ia models of \cite{wal12} to determine if variations in metallicity can explain this observed $UV-b$ dispersion. We find that metallicity variation can result in a $UV-b$ dispersion of 0.24 mag (larger than the observed 0.21 mag). Therefore, the increased dispersion at NUV compared to optical wavelengths is consistent with metallicity variations within the SN Ia sample, which manifest themselves at shorter wavelengths.

\subsection{Velocity shifts and the connection to SN Ia progenitors}
In Section \ref{results2}, we found that `normal' SNe Ia with broader than average light curves (higher stretch) have NUV features with systematically higher velocities than those with narrow light curves (lower stretch). This effect is most obvious in the \Caii\ H\&K and NUV $\lambda_2$ (3180 \AA, labelled in Fig. \ref{mean_highlow}) features, although a smaller effect is also present in the \SiII\ 4130 \AA\ and $\lambda_1$ (2920 \AA) features. In Fig. \ref{cahk_str_col}, we have plotted the best-fitting line to our low-$z$ data to show the correlation between \Caii\ H\&K velocity and stretch, which has a significance of 3.4-$\sigma$. However, it is unclear if instead of a correlation, we are seeing a bi-modality in the data, with SNe with low stretches ($s<1.01$) having lower \Caii\ H\&K velocities and those with higher stretches ($s>1.01$) having higher \Caii\ H\&K velocities. We independently confirm the presence of these trends using the spectral fitting code, \textsc{synapps} \citep{tho11}, where the model \Caii\ and \SiII\ velocities outputted are also found to show this trend of increasing velocity with stretch. \cite{fol12} did not identify this trend between \Caii\ H\&K velocity and stretch, probably due to the restrictive stretch cut of $s<1.05$ employed, which removes 7 high-stretch SNe Ia from their sample.

Similar trends of increasing velocity \citep{wel94, fis95, maz98, maz07} of NUV and optical spectral features with increasing light curve width (increasing luminosity) have also been identified in previous samples of maximum light, as well as nebular phase, spectra. In these studies, the trend was observed when a much broader sample of SNe Ia were studied, including sub-luminous 91bg-like SNe Ia \cite[][]{fil92, lei93} and over-luminous 91T-like SNe Ia \citep{fil92b}.  Instead our sample focusses solely on a cosmologically useful SN Ia sample, and it is within this narrower sample that a trend of increasing NUV velocities with increasing light curve width (stretch) is observed. 

A possible cause of this correlation between velocity and light curve width could be differences in the kinetic energies of the SNe: high luminosity (high stretch) SNe Ia have larger \nick\ masses, which may lead to larger kinetic energies, and hence higher velocities. Additionally, higher velocities lead to stronger line blanketing in the NUV (due to more overlapping of the spectral lines), which could cause a depression of the UV pseudo-emissions, as is seen for our high stretch sub-sample in Fig. \ref{mean_stretch2}. 
High-velocity components (from detached shells of material) have been identified in most SNe Ia with spectra obtained at phases earlier than one week before maximum light \citep{maz07, blon12}, which may influence the measurements of spectral velocities at these early times. Blending of high-velocity features with photospheric features could result in higher overall velocities being measured.  However, the level of persistence of these high-velocity features to maximum light epochs is unclear and in the case of the \Caii\ H\&K feature, can be confused with the \SiII\ feature which is present in the same wavelength region. Therefore, we have used the \textsc{synapps} spectral models to distinguish between these scenarios and show that \Caii\ H\&K velocity does correlate with stretch for our sample.

Features at NUV wavelengths, including the $\lambda_2$ feature, have been found using spectral synthesis modelling to be dominated by reverse fluorescence processes with the dominant species that contribute being \Mgii, \SiII, \SiiNF, along with some Fe-group elements \citep{sau08, wal12}. The $\lambda_1$ and $\lambda_2$ pseudo-emission features can be considered simply as regions of lower opacity, and will have bluer positions if the velocities and kinetic energies are higher.

The mean spectra split by host galaxy M$_{\rm stellar}$ and also the analysis of the individual features show a trend of increased \Caii\ H\&K velocity/blue-shifted $\lambda_2$ feature for the low M$_{\rm stellar}$ galaxies, or equivalently we see a lack of low-velocity/less blue-shifted events in low M$_{\rm stellar}$ galaxies, and lack of high-velocity/more blue-shifted events in high M$_{\rm stellar}$ galaxies, as claimed by \cite{fol12}. However, we also show that this relation is driven by our identified trend between velocity and stretch, and the well-known correlation that brighter (higher stretch) SNe Ia favour late-type (low mass) galaxies \citep{ham96, ham00, sul06}.

Higher velocity spectral features have also been found to be more common in SNe Ia that display outflowing material in high-resolution spectra, suggestive of a single-degenerate progenitor channel \citep{ste11, fol12c}. This suggests that we can distinguish different progenitor scenarios using spectral feature velocities. In this paper, we find that higher stretch (higher luminosity) SNe Ia have, on average, higher velocities, and these SNe are more common in low M$_{\rm stellar}$ galaxies. Following this argument to its logical conclusion, we suggest that SNe Ia in low M$_{\rm stellar}$ galaxies (predominantly young stellar populations), should result from the single-degenerate progenitor channel. Conversely, the lack of evidence of outflowing material in SNe Ia with low velocities and that form more commonly in high M$_{\rm stellar}$ galaxies (predominantly older stellar populations), suggest that these progenitor systems may be linked to the double-degenerate progenitor channel.

\subsection{Cosmological implications}
The low-$z$ mean spectra can be split by the value of the Hubble residuals, calculated from the light curve fits to investigate the effect of NUV spectral variations on the use of SNe Ia for cosmology. We find no noticeable difference in the mean flux when the sample is split into positive and negative Hubble residuals. We also observe no correlation between the stretch- and colour-corrected, as well as only colour-corrected, residuals and the velocities/wavelengths of the NUV spectral features. This suggests that the observed NUV spectral variations do not directly translate into a cosmological bias. However, our observed NUV redshift evolution could have a secondary impact on NUV-optical colours, which enter some distance estimation techniques \citep[e.g.][]{con08}.

\section{Conclusions}

We have performed a detailed study of 32 near-UV (NUV) spectra of low-$z$ ($0.001<z<0.08$) SNe Ia using \textit{HST}+STIS. These spectra are complemented by $gri$-band light curves for the entire sample from the LT and LCOGT robotic telescopes. We compared our low-$z$ sample to that of the intermediate sample of E08 to test for evolution with redshift, as well as investigating correlations between spectral, light curve and host galaxy properties within the low-$z$ sample. Our principal conclusions are:
\begin{enumerate}
\item{We observe an evolution with redshift of the near-UV continuum at the 3-$\sigma$ level, with low-$z$ SNe Ia having depressed near-ultraviolet flux compared with the intermediate-$z$ sample of E08 (Fig. \ref{mean_highlow}). This is consistent with measurements of the $UV-b$ colour, where the low-$z$ sample has a 0.18 (3.6-$\sigma$) mag redder colour than the intermediate-$z$ sample.} 
\item{We identify an increased dispersion at near-UV compared to optical wavelengths in our low-$z$ SN Ia sample, as was found by E08 for an intermediate-$z$ sample (Fig. \ref{mean_highlow}).}
\item{Using the SN Ia spectral synthesis models of \cite{wal12}, we have shown that this redshift evolution can be explained by a decrease in metallicity with increasing redshift of $\sim$0.4 dex. The increased dispersion seen at near-UV compared to optical wavelengths can be explained using the same models by metallicity variations with the SN Ia sample. The observed redshift evolution is in agreement with galaxy evolution studies show a $\sim$0.2 dex decrease in metallicity over this redshift range.}
\item{We find that SNe Ia with broader light curves (higher stretches) have systematically higher expansion velocities, in particular seen in the \Caii\ H\&K feature (Fig. \ref{cahk_str_over}, Fig. \ref{cahk_str_col}), as well as blue-shifted NUV spectral features such as $\lambda_2$ (Fig. \ref{lam2_pos_colstr}) . No correlation between velocity and $B-V$ colour is found.}
\item{We show that SNe Ia in low M$_{\rm stellar}$ host galaxies have high NUV spectral feature velocities, as claimed by \cite{fol12}. However, we attribute it to our velocity versus light curve width (stretch) correlation, and the well-known result that higher luminosity, broader light curve SNe Ia are preferentially found in low M$_{\rm stellar}$ galaxies (Fig. \ref{cahk_mass_vel}). }
\item{We construct mean spectra of our low-$z$ sample, split by Hubble residual, and find little variation in the mean spectra, indicating that NUV variations do not result in a clear SN Ia cosmology bias (Fig. \ref{hub_mean}).}
\end{enumerate}

Our results underline the critical importance of the UV region for SN Ia studies. Given the qualitative success of the SN Ia models in reproducing the trends in our data, a more quantitative analysis using detailed spectra can be performed, to obtain important parameters such as the density structure and elemental abundances of the individual SN events. This will enable us to put tighter constraints on the progenitor systems of SNe Ia, and how their progenitor channels may vary with host galaxy properties.

\section{Acknowledgements}

MS acknowledges support from the Royal Society. AG-Y and MS
acknowledge support from the Weizmann-UK ``making connections''
programme. A.G. further acknowledges support by the ISF, BSF, a Minerva grant, the ARCHES
award, and the Lord Sieff of Brompton Fund.  DAH and BD are supported by the Las Cumbres Observatory Global Telescope Network. MMK acknowledges generous support from the Hubble Fellowship and
Carnegie-Princeton Fellowship. EOO is incumbent of
the Arye Dissentshik career development chair and
is grateful to support by
a grant from the Israeli Ministry of Science.

Based on observations made with the NASA/ESA Hubble Space Telescope,
obtained at the Space Telescope Science Institute, which is operated
by the Association of Universities for Research in Astronomy, Inc.,
under NASA contract NAS 5-26555. These observations are associated
with programs 11721 and 12298. The Liverpool Telescope is operated on
the island of La Palma by Liverpool John Moores University in the
Spanish Observatorio del Roque de los Muchachos of the Instituto de
Astrofisica de Canarias with financial support from the UK Science and
Technology Facilities Council.  Observations were obtained with the
Samuel Oschin Telescope at the Palomar Observatory as part of the
Palomar Transient factory project, a scientific collaboration between
the California Institute of Technology, Columbia Unversity, La Cumbres
Observatory, the Lawrence Berkeley National Laboratory, the National
Energy Research Scientific Computing Center, the University of Oxford,
and the Weizmann Institute of Science. The William Herschel Telescope
is operated on the island of La Palma by the Isaac Newton Group in the
Spanish Observatorio del Roque de los Muchachos of the Instituto de
Astrofísica de Canarias. Some of the data were obtained with the W. M.
Keck Observatory, which is operated as a scientific partnership among
the California Institute of Technology, the University of California
and the National Aeronautics and Space Administration. These
observations were made possible by the generous financial support of
the W. M. Keck Foundation. This paper uses observations obtained with facilities of
the Las Cumbres Observatory Global Telescope. The
Byrne Observatory at Sedgwick (BOS) is operated by
the Las Cumbres Observatory Global Telescope Network
and is located at the Sedgwick Reserve, a part of the
University of California Natural Reserve System. This research has made use of the NASA/IPAC Extragalactic Database (NED) which is operated by the Jet Propulsion Laboratory, California Institute of Technology, under contract with the National Aeronautics and Space Administration. 

This publication has been made possible by the participation of more
than 10,000 volunteers in the Galaxy Zoo: Supernovae project,
\texttt{http://supernova.galaxyzoo.org/authors}.

\appendix
\section{Wavelength and velocity measurements}

\begin{table*}
 \caption{Velocity measurements for \Caii\ H\&K and \SiII\ 4130 \AA, along with wavelength measurements for the $\lambda_2$ and $\lambda_1$ features. Both the measured values and the phase-corrected 0 d measurements are given. }
 \begin{tabular}{@{}lccccccccccccccccccccccccccccccc}
  \hline
  \hline
SN name&Measured &Phase-corrected& Measured &Phase-corrected & Measured&Phase-corrected&Measured&Phase-corrected\\ 
& \Caii\ ($10^3$ \kms)&\Caii\ ($10^3$ \kms)&\SiII\ ($10^3$ \kms)&\SiII\ ($10^3$ \kms)&$\lambda_1$ (\AA)  &$\lambda_1$ (\AA)&$\lambda_2$ (\AA)&$\lambda_2$ (\AA)\\ 
\hline
PTF09dlc   &     16.99$\pm$0.17  &    17.78$\pm$0.67	   &       	9.29$\pm$0.05     &      9.52$\pm$0.17& --	  &--     &					--	   &	--						    \\
PTF09dnl   &     16.81$\pm$0.15  &    17.17$\pm$0.33    &         10.09$\pm$0.03   &        10.20$\pm$0.08&   2974$\pm$3     &    2976$\pm$5       &  3155.5$\pm$3.4        &   3147.3$\pm$5.4    \\ 
PTF09dnp   &      --				  &   			--	    &         10.98$\pm$0.05   &         11.51$\pm$0.35&    2952$\pm$3     &    2962$\pm$16         &   3168.1$\pm$3.3       &   3132.2$\pm$15.6    \\
PTF09fox   &     14.40$\pm$0.04   &   	15.14$\pm$0.62   &          10.14$\pm$0.09  &         10.38$\pm$0.18&    2963$\pm$20   &      2967$\pm$21    &   3160.2$\pm$4.8       &    3143.9$\pm$8.6    \\
PTF09foz   &     14.28$\pm$0.11  &    15.07$\pm$0.67    &         8.55$\pm$0.06    &          8.81$\pm$0.18&  2976$\pm$5      &   2982$\pm$9        & 3170.3$\pm$4.6         &  3152.8$\pm$8.9    \\ 
PTF10bjs   &     16.47$\pm$0.22  &    17.02$\pm$0.50    &         12.69$\pm$0.06   &        12.87$\pm$0.13&    2930$\pm$5    &     2933$\pm$7      &   3125.2$\pm$4.7       &    3113.1$\pm$7.1    \\ 
PTF10fps   &    		--			  &   			--		    &         7.40$\pm$0.05    &           8.37$\pm$0.62&   2927$\pm$4      &   2946$\pm$29          &  3181.9$\pm$5.3        &  3117.2$\pm$28.1    \\
PTF10hdv   &     17.03$\pm$0.18  &    17.94$\pm$0.78    &         10.13 $\pm$0.04   &        10.42$\pm$0.20&    2962$\pm$5    &     2967$\pm$10     &  3136.9$\pm$10.5       &    3117.1$\pm$13.2    \\ 
PTF10hmv   &     15.02$\pm$0.11  &    15.72$\pm$0.59    &         9.29  $\pm$0.05    &       9.51$\pm$ 0.15&  2970$\pm$4      &   2974$\pm$8        & 3109.8$\pm$4.6         &  3094.8$\pm$8.3    \\ 
PTF10icb   &     12.61$\pm$0.03   &   	12.85$\pm$0.20   &          9.11 $\pm$0.02   &        9.19$\pm$0.05& 2953$\pm$3       &  2954$\pm$4         &3154.2$\pm$2.3          & 3148.8$\pm$3.1    \\ 
PTF10mwb   &     13.09$\pm$0.03   &   	12.99$\pm$0.09   &          8.21 $\pm$0.04   &        8.18$\pm$0.05& 2949$\pm$5       &  2948$\pm$5         &3174.4$\pm$3.2          & 3176.2$\pm$3.5    \\ 
PTF10ndc   &     14.20$\pm$0.12  &   	 15.84$\pm$1.37   &         10.18$\pm$0.15   &        10.71$\pm$0.38&   2930$\pm$25    &  2940$\pm$30        &  3171.2$\pm$4.9        &  3135.3$\pm$18.3 \\ 
PTF10nlg   &     15.20$\pm$0.17  &   	 16.93$\pm$1.45   &         9.32 $\pm$0.09    &       9.89$\pm$0.37&  2966$\pm$5      &  2976$\pm$18        &3165.1$\pm$4.3          &3127.9$\pm$16.6    \\ 
PTF10qjl   &     14.88$\pm$0.19  &   	 16.55$\pm$1.40   &         9.33 $\pm$0.06    &       9.88$\pm$0.35&  2931$\pm$15     &   2941$\pm$22       & 3136.7$\pm$3.1         & 3100.6$\pm$16.1    \\ 
PTF10qjq   &     11.87$\pm$0.08   &   	12.86$\pm$0.83   &          10.02$\pm$0.05  &         10.34$\pm$0.21&    2945$\pm$5    &    2951$\pm$11      &  3174.1$\pm$2.2        &  3152.4$\pm$9.6   \\ 
PTF10tce   &     16.02$\pm$0.05   &   	17.00$\pm$0.81    &         10.94$\pm$0.08   &        11.25$\pm$0.22&    2931$\pm$3    &    2937$\pm$10      &  3160.6$\pm$5.3        &  3139.7$\pm$10.4    \\
PTF10wnm   &     12.90$\pm$0.07   &   	14.04$\pm$0.96    &         10.06$\pm$0.1   &        10.442$\pm$0.27&    2947$\pm$37   &     2954$\pm$12     &   3164.1$\pm$2.1       &   3139.0$\pm$10.9    \\
PTF10wof   &     13.83$\pm$0.06   &   	15.49$\pm$1.39   &         9.35$\pm$0.06    &       9.89$\pm$0.35&  2989$\pm$5      &  2999$\pm$17        &3170.2$\pm$5.4         &    3133.7$\pm$16.4   \\
PTF10xyt   &     15.01$\pm$0.08   &   	15.90$\pm$0.74    &         10.86$\pm$0.1   &        11.157$\pm$0.25&    2933$\pm$65   &     2938$\pm$11     &   3155.1$\pm$6.7       &   3135.6$\pm$10.9    \\
PTF10ygu   &     19.09$\pm$0.08   &   	18.99$\pm$0.12    &      			--			   &     	--				&   2928$\pm$10    &    2927$\pm$10      &  3155.3$\pm$3.9        &  3157.1 $\pm$4.1   \\
PTF10yux   &     16.05$\pm$0.13  &   	 18.04$\pm$1.66   &         6.97$\pm$0.14   &        7.61$\pm$0.44&   2923$\pm$3     &   2935$\pm$20       &   	--    &        		--			\\
PTF10zdk   &     16.95$\pm$0.14  &   	 16.22$\pm$0.63    &         8.30$\pm$0.04    &       8.06$\pm$0.16&  2913$\pm$6      &  2909$\pm$9         & 3115.4$\pm$ 5.4         & 3131.0$\pm$9.0    \\
PTF11kly    &    11.93$\pm$0.04   &  	 12.72$\pm$0.66     &        8.54$\pm$0.13    &       8.79$\pm$0.21 & 2954$\pm$1      &  2959$\pm$8         & 3186.9$\pm$ 1.2         & 3169.8$\pm$7.4    \\
SN2009le   &    16.08$\pm$0.12   &   16.16$\pm$0.14       &        11.83$\pm$0.1    &       11.86$\pm$0.11 &   2948$\pm$31   &    2948$\pm$3      &  3148.2$\pm$  3.3     	  &    3146.3$\pm$3.1 	  \\
SN2010ju   &    16.95$\pm$0.07    &  	18.48$\pm$1.28    &        9.99$\pm$0.01     &      10.48$\pm$0.32 &  2942$\pm$5     &   2951$\pm$16      &   3119.1$\pm$9.4      &    3085.6$\pm$16.9   \\
SN2011by   &    12.64$\pm$0.06    &  	12.56$\pm$0.09      &        9.09$\pm$0.01     &      9.06$\pm$0.02 &    2958$\pm$2   &     2958$\pm$2     &  3166.5$\pm$1.1       &    3167.9$\pm$1.3   \\
SN2011ek   &    12.80$\pm$0.10    &  	13.85$\pm$0.88     &        9.33$\pm$0.01     &      9.67$\pm$0.22 &	--    & 	-- &    	--   	  &     		-- 			 \\
\hline	
 \end{tabular}
  \label{vel_spec}
\end{table*}

\end{document}